\documentclass[fleqn,usenatbib]{mnras}

\usepackage{newtxtext,newtxmath}
\usepackage{dcolumn}
\newcolumntype{d}[1]{D{.}{.}{#1}}

\usepackage[T1]{fontenc}
\usepackage{ae,aecompl}

\usepackage{graphicx}	
\usepackage{amsmath}	
\usepackage{amssymb}	
\usepackage{booktabs}
\usepackage{bm}
\usepackage{multirow}
\usepackage{rotfloat}
\usepackage[tableposition=top]{caption}
\usepackage{pdflscape}

\newcommand{\lapprox }{{\lower0.8ex\hbox{$\buildrel <\over\sim$}}}
\newcommand{\gapprox }{{\lower0.8ex\hbox{$\buildrel >\over\sim$}}}

\newcommand{\given}{\,|\,}
\newcommand{\dd}{\mathrm{d}}

\newcommand{\degree}{\ifmmode {^\circ}\else$^\circ$\ \fi}
\newcommand{\amin}{\ifmmode {^{\prime}\ }\else$^{\prime}$\fi}
\newcommand{\asec}{\ifmmode {^{\prime\prime}}\else$^{\prime\prime}$\fi}

\newcommand{\Msun}{\ifmmode {M_{\odot}}\else${M_{\odot}}$\fi}
\newcommand{\Rsun}{\ifmmode {R_{\odot}}\else${R_{\odot}}$\fi}
\newcommand{\matr}[1]{\bm{#1}}

\title[Wide Binaries in TGAS]{Wide Binaries in Tycho-{\it Gaia}: Search Method and the Distribution of Orbital Separations}

\author[Andrews, Chanam\'e, \& Ag\"ueros]{Jeff J. Andrews$^{1,2}$\thanks{Contact e-mail: \href{mailto:andrews@physics.uoc.gr}{andrews@physics.uoc.gr}}, Julio Chanam\'{e}$^{3,4}$, Marcel A.~Ag\"{u}eros$^5$ \\
$^1$ Foundation for Research and Technology - Hellas, IESL, Voutes, 71110 Heraklion, Greece \\
$^2$ Physics Department \& Institute of Theoretical \& Computational Physics, University of Crete, 71003 Heraklion, Crete, Greece \\
$^3$ Instituto de Astrof\'isica, Pontificia Universidad Cat\'olica de Chile, Av. Vicu\~na Mackenna 4860, 782-0436 Macul, Santiago, Chile \\
$^4$ Millennium Institute of Astrophysics, Santiago, Chile \\
$^5$ Department of Astronomy, Columbia University, 550 West 120th Street, New York, NY 10027, USA
} 

\date{Accepted XXX. Received YYY; in original form ZZZ}

\pubyear{2017}

\begin{document}
\label{firstpage}
\pagerange{\pageref{firstpage}--\pageref{lastpage}}
\maketitle

\begin{abstract}
We mine the Tycho-{\it Gaia} astrometric solution (TGAS) catalog for wide stellar binaries by matching positions, proper motions, and astrometric parallaxes. We separate genuine binaries from unassociated stellar pairs through a Bayesian formulation that includes correlated uncertainties in the proper motions and parallaxes. Rather than relying on assumptions about the structure of the Galaxy, we calculate Bayesian priors and likelihoods based on the nature of Keplerian orbits and the TGAS catalog itself. We calibrate our method using radial velocity measurements and obtain 6196 high-confidence candidate wide binaries with projected separations $s\lesssim1$ pc. The normalization of this distribution suggests that at least 0.6\% of TGAS stars have an associated, distant TGAS companion in a wide binary. We demonstrate that {\it Gaia}'s astrometry is precise enough that it can detect projected orbital velocities in wide binaries with orbital periods as large as 10$^6$ yr. For pairs with $s\ \lapprox\ 4\times10^4$~AU, characterization of random alignments indicate our contamination to be $\approx$5\%. For $s \lesssim 5\times10^3$~AU, our distribution is consistent with \"{O}pik's Law. At larger separations, the distribution is steeper and consistent with a power-law $P(s)\propto s^{-1.6}$; there is no evidence in our data of any bimodality in this distribution for $s \lesssim$ 1 pc. Using radial velocities, we demonstrate that at large separations, i.e., of order $s \sim$ 1 pc and beyond, any potential sample of genuine wide binaries in TGAS cannot be easily distinguished from ionized former wide binaries, moving groups, or contamination from randomly aligned stars.
\end{abstract}




\section{Introduction}
\label{sec:intro}
Roughly half of all main-sequence stars are found in binary systems with orbital separations extending from $\sim$10~\Rsun\ to $\sim$pc \citep{duquennoy91,fischer92,raghavan10,duchene2013}. Unresolved binaries at the smallest separations are typically found by identifying radial velocity (RV) variations indicative of orbital motion, while binaries at intermediate separations can be identified by observing the stars' motions through their orbits over many years \citep{raghavan10}, or by identifying in decades of archival data dips in a star's luminosity caused by eclipses due to an unseen stellar companion \citep{rodriguez16}. At the largest separations, stars are typically associated by watching them move through the solar neighborhood in the same direction and at the same rate; these common proper motion pairs have separations from 10$^2$ to $\geq$10$^4$~AU \citep{luyten71}.

The orbits of wide binaries may represent a significant fraction of the size of the clusters within which most stars are born \citep{lada03}. Such widely separated binaries are likely to form during the dissolution of a single-age stellar cluster \citep{kouwenhoven10, moeckel11} or from the evolution of higher-order systems \citep{reipurth12,elliott16}. Since these pairs may be dynamically removed or slowly separated from the cluster \citep{theuns92,fujii12}, the population of wide binaries can probe the physics of cluster dissolution \citep{kroupa01, kouwenhoven10}. Additionally, triples or higher order systems formed via the fragmentation of circumbinary disks \citep{bonnell94b} may evolve dynamically into independent binaries \citep{bate02,moeckel10} or stable, hierarchical triple systems \citep{reipurth12}. The observed distribution of orbital periods is therefore the result of the physics of formation and of subsequent dynamical reprocessing \citep{kroupa95, jiang10, moeckel11}.

Beyond $\approx$10$^3$~AU, the components of wide binaries  are not expected to interact significantly over their lifetimes. These pairs can be considered mini-open clusters \citep{binney87,soderblom10} and are uniquely powerful for answering questions that cannot be addressed with other stellar systems. For example, wide binaries consisting of white dwarfs, subdwarfs, or M-dwarfs with FGK main sequence companions have been used to calibrate metallicity scales and age-rotation relations \citep{lepine07b, rojas-ayala10, garces11, chaname2012, li14}. Other white dwarf-main sequence pairs and double white dwarfs have been used to determine the initial-to-final mass relation for white dwarfs \citep{finley97, zhao12, andrews15}. At the widest separations, binaries are affected by interactions with other stars, giant molecular clouds, and the Galactic tide, each of which acts to disrupt them stochastically \citep{weinberg87, jiang10}. These binaries have been used as probes of the gravitational potential of their host, including the contribution of dark objects and of dark matter in general \citep{bahcall85, yoo04, hernandez08, quinn09b, monroy-rodriguez14, penarrubia16}.

We define a {\it bona fide} wide binary as a system of two stars whose properties are consistent with being gravitationally bound to each other. We do not claim that the components of wide binaries presented in this work are demonstrably orbiting their common center of mass, as that would at least require measurement of the accelerations of the components or of their orbit on the plane of the sky \citep[e.g.,][]{tokovinin16}. Instead, we regard a pair as a genuine binary when, to the limits of the  data, their phase-space coordinates are consistent with the hypothesis that they are gravitationally bound to each other. 

This definition becomes crucial when considering how wide one can go in a search for wide binaries. Within the Milky Way, a pair of stars cannot remain gravitationally bound for long beyond the  separation at which they become sensitive to the Galactic tidal field. This tidal limit (or Jacobi radius) is $\approx$2 pc for $\approx$\Msun\ stars within the solar neighborhood \citep{yoo04,jiang10}. Pairs wider than this limit do exist: they may be wide binaries that have been ionized, members of moving groups, the tidal remains of larger structures, etc. However, as we demonstrate later in this work, they overlap in observational space with a sizable population of (false) pairs that occur simply by chance. Given the large variety of astronomical problems that can benefit from the assembly of reliable samples of wide binaries, we argue that pairs beyond the Galactic tidal limit should not be confused with genuine wide binaries, and we limit our search to pairs with projected separations $\lesssim$1 pc.

The exact form of the distribution of {\it bona fide} wide binary separations can help to constrain their formation scenarios. Although some groups find that this distribution is flat in log space out to the largest separations \citep[following what is known as \"{O}pik's Law; e.g.,][]{opik24, poveda07, allen14}, the current consensus favors a distribution that decreases at larger separations \citep{chaname04, lepine07, tokovinin12, halbwachs17}. As of yet, however, only a few attempts have been made to link directly the observed orbital separation distribution with models: \citet{kouwenhoven10} and \citet{moeckel11} have both compared $N$-body simulations of cluster dissolution to the observed orbital separation distribution. One reason for this scarcity of comparisons between data and simulations is that the observed samples are typically small. Identifying large numbers of wide binaries is difficult, and samples often suffer from significant and hard-to-characterize observational biases.

Pioneering work identifying wide binaries focused on small, specialized catalogs of stars \citep{luyten71, wasserman91, garnavich91, gould95}. More recent efforts have employed methods of varying complexity to systematically search larger, astrometric catalogs such as those produced by the New Luyten Two-Tenths survey \citep[NLTT;][]{luyten79} and the revised NLTT \citep[rNLTT;][]{gould03, salim03}, e.g., \citet{allen00, chaname04}, the Sloan Digital Sky Survey \citep[SDSS;][]{york00}, e.g., \citet{sesar08, quinn09a, dhital10, longhitano10, andrews12, baxter14, andrews15, dhital15}, the L{\'e}pine-Shara Proper Motion-North catalog \citep[LSPM;][]{lepine05}, e.g., \citet{lepine07}, and {\it Hipparcos} \citep{hip}, e.g., \citet{shaya11, tokovinin12}. These searches typically fall in one of two categories: there are those where stars are matched using only position (for instance using a two-point correlation function) and those where proper motion is included as well (to identify common proper motion pairs), corresponding to matching two or four dimensions, respectively, of the six-dimensional phase space.\footnote{Note that several groups have included photometric distance estimates in their matching algorithm. Although not used within the matching algorithm itself, there have also been efforts to include RVs \citep[e.g.,][]{latham84}, astrometric parallaxes \citep[e.g.,][]{lepine07}, or both \citep{halbwachs17} as extra phase-space dimensions to confirm the binary nature of wide pairs.} 

\cite{shaya11} performed the first large-scale search for binaries that included {\it Hipparcos} parallaxes in a matching algorithm. The additional phase-space dimension adds a powerful constraint for separating genuine binaries from randomly aligned stellar pairs. However, because of its non-linear dependence on distance and non-trivial prior probabilities, the inclusion of parallax also adds a significant degree of complexity to the matching algorithm. Nevertheless, using a Bayesian formalism, these authors identify over 1000 binaries, including several high-probability pairs with projected separations of $\sim$pc. 

The first data release from the {\it Gaia} satellite \citep{gaia_DR_1, gaia_DR1_2, lindegren16} includes $2\times10^6$ Tycho-2 stars \citep{hog00}, for which a Tycho-{\it Gaia} astrometric solution \citep[TGAS;][]{michalik15} is provided. The TGAS catalog is nearly complete to $V=11$ mag, with stars out to $V\approx12$. Each of these stars has positions (with uncertainties of $\approx$0.4 mas), proper motions (uncertainties $\approx$1.3 mas yr$^{-1}$), and parallaxes \citep[uncertainties $\approx$0.3 mas, although a systematic bias of $\approx$0.25 mas has been identified in {\it Gaia} parallaxes;][]{stassun16}. 

\citet{oelkers17} and \citet{oh2016} performed searches for wide binaries and moving groups in the TGAS catalog, including for subsets with separations $>$1 pc. The initial \citet{oelkers17} catalog includes 8516 candidate pairs in which both stars are in TGAS. Although these authors use a sophisticated Galactic model to determine the random alignment likelihood for any particular wide binary, they do not calculate a corresponding likelihood that any binary could be a genuine pair. Furthermore, while their method combines proper motions, parallaxes, and their associated uncertainties to identify real pairs, it neglects correlations between the proper motion and parallax uncertainties, which can be considerable. 

\citet{oh2016} use an improved method to identify co-moving stars. These authors calculate the likelihood that a particular pair is a genuinely associated binary and that it is a random alignment of unassociated stars. In calculating these two likelihoods, these authors assume a model for the stellar components of the Milky Way. 

We describe a Bayesian method for identifying wide binaries that combines positions, proper motions, astrometric parallaxes and their correlated uncertainties. Similar to \citet{oh2016}, we calculate two likelihoods for each stellar pair. However, our calculations use a data-based approach and are independent of any assumptions about the structure of the Milky Way.\footnote{Furthermore, \citet{oh2016} select only those pairs with a parallax signal-to-noise (S/N) ratio $>$8, whereas we search the entire TGAS catalog.} We separate genuine binaries from randomly aligned pairs using a posterior probability that combines these two likelihoods with prior probabilities based on the local stellar density. We further improve on previous work by verifying and calibrating our algorithm using two independent methods. First, using our algorithm, we identify and characterize a set of random alignments by matching the catalog to a shifted version of itself, which produces a sample containing only random alignments. Second, we use the subset of our pairs with radial velocity (RV) measurements to determine the rate of contamination due to randomly aligned stars. Our discussion is focused on identifying pairs within the TGAS catalog, but our method is designed to be adaptable to any stellar catalog with five-dimensional astrometry and is scalable to the expanded catalogs expected from future {\it Gaia} data releases. 

In Section~\ref{sec:challenges}, we outline the challenges posed by the size of the TGAS catalog and the quality of {\it Gaia} data. Our statistical method for combining positions, proper motions, and parallaxes is described in Section~\ref{sec:method}.  Before applying it to the full TGAS catalog, we describe in Section~\ref{sec:contamination} our efforts to predict the characteristics of the contamination due to false binaries. In Section~\ref{sec:gaia}, we assemble our candidate TGAS pairs, calibrate our method to set a probability threshold for identifying real binaries, and estimate our contamination by false binaries. We present the resulting catalog of wide binaries, and use a low-contamination subset to characterize the projected separation distribution in Section~\ref{sec:results}. In Section~\ref{sec:discussion}, we compare our sample to pre-TGAS catalogs of wide binaries, discuss the impact of our assumptions on our search, address the question of whether we can reliably identify pairs at separations $\gapprox$1 pc, and provide some guidance to users of our catalog. We conclude in Section~\ref{sec:conclusions} with a few thoughts about identifying wide binaries in future {\it Gaia} catalogs. Appendix~\ref{sec:rNLTT} is a test of our method on the subset of the rNLTT catalog with {\it Hipparcos} parallaxes; we compare the resulting catalog with a previously identified catalog of rNLTT wide binaries.

\section{Challenges}
\label{sec:challenges}

{\it Hipparcos} provided the only data comparable to the TGAS data, for $\approx$10$^5$ stars. Since the number of possible pairs of a catalog of size $N$ is the arithmetic mean $N(N-1)/2$, identifying binaries in {\it Hipparcos} data implies the examination of $\approx$$5\times10^9$ individual pairs of stars. While the TGAS catalog is only 20 times larger than the {\it Hipparcos} catalog, the total number of pairs it contains is $\approx$$2\times10^{12}$, a factor of 500 increase, and the final {\it Gaia} catalog will contain $\approx$$5\times10^{17}$ individual pairs. Of course, not every stellar pair need be checked (for instance, one can select a maximum angular search radius around each star), but the $\approx$$N^2$ scaling with catalog size is likely unavoidable. Clearly the algorithms developed to search previous astrometric catalogs, even those designed to include {\it Hipparcos} parallaxes, need to be rethought. 

\begin{figure}
\centerline{\includegraphics[width=1.07\columnwidth]{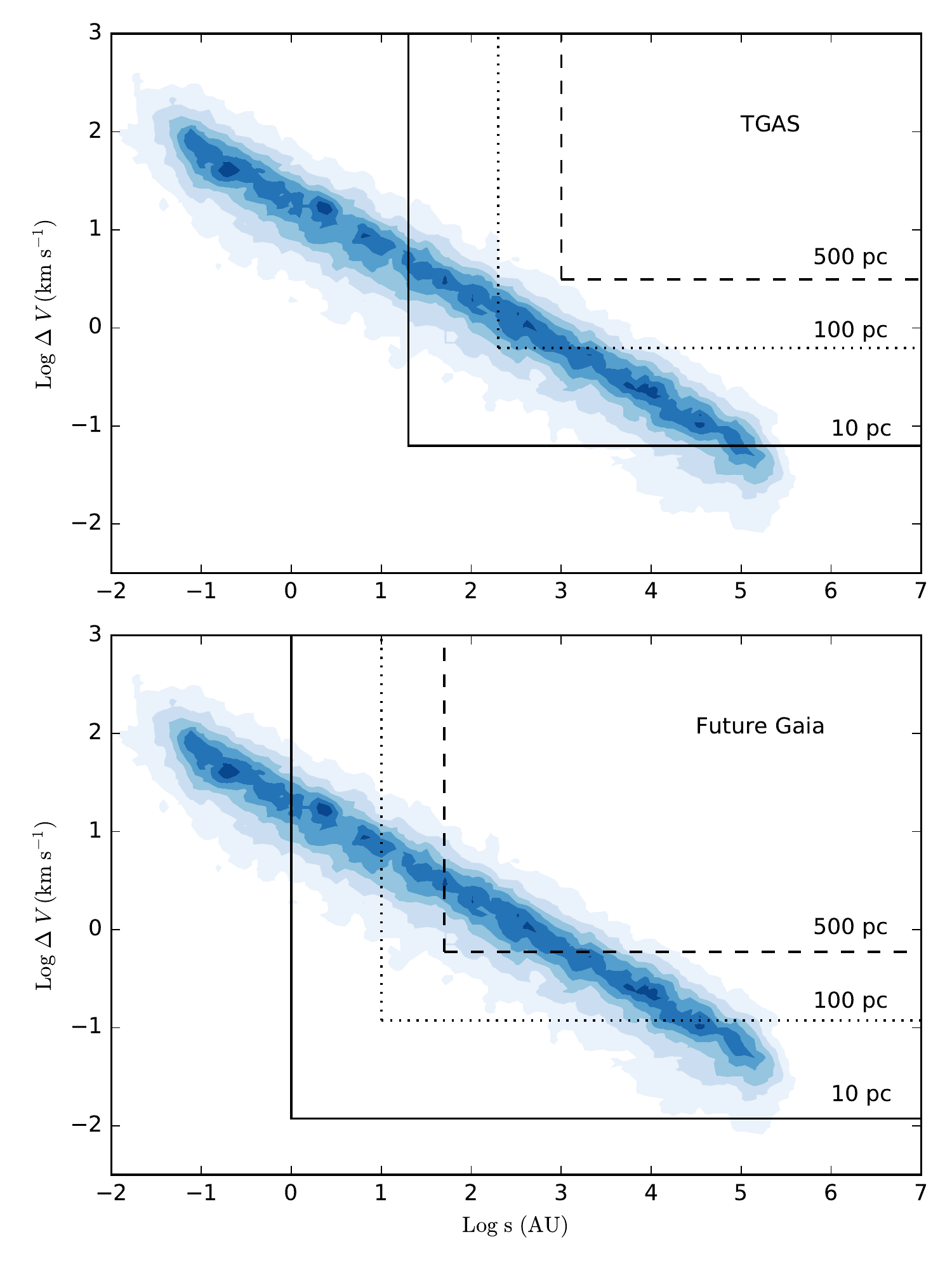}}
\caption{The kernel density estimate representation of  10$^4$ randomly produced binaries in $\log s- \log \Delta V$ space. Binaries with the largest separations ($\approx$1 pc) should have nearly identical proper motions. For smaller projected separations, the tangential velocities may differ by as much as tens of km s$^{-1}$. The top panel shows the approximate current {\it Gaia} sensitivity of 1.3 mas~yr$^{-1}$ in proper motion and 2\asec\ in angular separation for three different distances. Differential velocities due to orbital motions can be ignored for binaries outside the boxed regions. However, {\it Gaia} can measure the $\Delta V$ for stars in binaries within those regions, so any identification algorithm needs to account for this $\Delta V$. The bottom panel shows that by the final {\it Gaia} data release, $\Delta V$ should be detectable for a much larger number of binaries. }
\label{fig:gaia_binary_limits}
\end{figure}

Furthermore, {\it Gaia}'s superb astrometric precision creates new challenges for identifying wide binaries. Given this precision, the traditional approach of selecting pairs of stars with identical proper motions will miss a large number of true binaries: {\it Gaia} is sensitive to the differential orbital velocities of many wide binaries. 

To illustrate, we generate 10$^4$ binaries with separations $a$ randomly drawn from a flat distribution in log space\footnote{Although evidence has indicated that the distribution of wide binary separations follows a power-law with a somewhat steeper index than $-$1 \citep{chaname04,lepine07}, adopting this flat distribution  more clearly demonstrates the region in $s-\Delta V$ space populated by binaries.} and eccentricities $e$ drawn from a thermal distribution:
\begin{eqnarray}
P(a) \propto a^{-1} &;& a \in [10\ R_{\odot}:1\ {\rm pc}] \\
P(e) \propto e\ \ \ &;& e \in [0:1).
\end{eqnarray}

We further adopt random binary orientation angles to determine the distribution of the projected separations $s$ and of the projected differences in orbital velocities $\Delta V$. 

Figure \ref{fig:gaia_binary_limits} compares the distribution of our binaries in $\log s-\log \Delta V$ space to {\it Gaia}'s astrometric limits. The TGAS data can be used to identify pairs of stars with angular separations as small as $\approx$2\asec\ and proper motions as small as $\approx$1.3~mas yr$^{-1}$. We apply these limits to pairs at three representative distances, 10, 100, and 500 pc, which allow us to translate the angular units observed by {\it Gaia} into physical units. In the top panel of Figure~\ref{fig:gaia_binary_limits}, we show the corresponding areas in $\log s-\log \Delta V$ space for which {\it Gaia} can measure $\Delta V$, the difference between the two velocity components of a binary due to orbital motion. 

By the final {\it Gaia} data release, the astrometry will be substantially more precise: on-ground processing may allow identification of stellar pairs with separations as small as 0.1\asec\ \citep{harrison11}, and proper motions may reach a precision of $\approx$0.25 mas yr$^{-1}$. Using these limits, we generate the rectangular areas in the bottom panel of Figure~\ref{fig:gaia_binary_limits}. For most of wide binaries within 100~pc, and for a substantial fraction of those within 500 pc, $\Delta V$ must be taken into account.

Figure \ref{fig:gaia_binary_limits} also highlights one of the primary motivations for characterizing the population of wide binaries: a substantial fraction ($\approx$50\%, assuming {\"O}pik's Law) have orbital separations $>$10$^2$ AU. Those binaries are too widely separated for a full orbit to be observed, and the only currently robust method to associate these pairs involves matching their position in phase space. Previous searches for such pairs were limited to nearby stars with large proper motions. With {\it Gaia}, we now have the ideal data to further explore this region of binary parameter space with significantly larger samples. 

Finally, {\it Gaia} presents an additional, unprecedented opportunity to access all six dimensions of phase space. Its final RV catalog will achieve a precision of $\approx$10 km~s$^{-1}$,  enough to aid substantially in separating genuine binaries from randomly aligned pairs. To our knowledge, RVs have been used exclusively in the past to check the binarity of wide binary candidates \citep[e.g.,][]{latham84, halbwachs17}; no RV-related algorithms have yet been included in common proper motion pair searches. For the first time, large numbers of binaries will be identified by their full three-dimensional positions and velocities.

\begin{figure}
\begin{center}
\includegraphics[width=1.05\columnwidth]{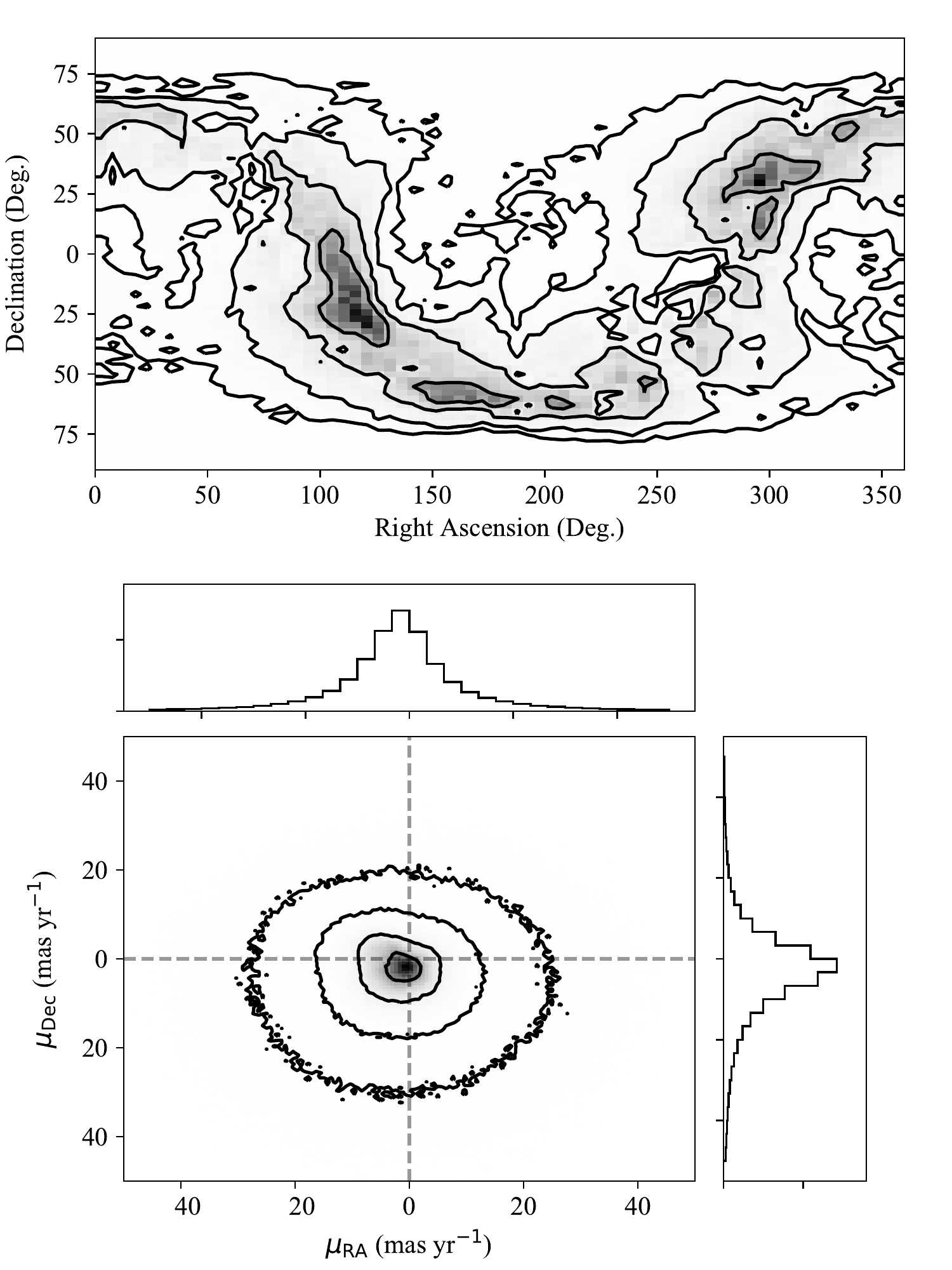}
\caption{ The Galactic Plane can be clearly identified in the positional distribution of TGAS objects in the top panel. The bottom panels shows that the majority of TGAS stars have proper motions $\lapprox$20 mas yr$^{-1}$. However, the tail of the proper motion distribution extends to several 10$^2$ mas yr$^{-1}$. }
\label{fig:tycho-2_pos_mu}
\end{center}
\end{figure}

\section{Statistical Method}
\label{sec:method}
Figure \ref{fig:tycho-2_pos_mu} shows the positional (top panel) and proper motion (bottom panel) distributions of stars in the TGAS catalog. The density of stars is clearly a function of position and proper motion, and the probability of identifying two randomly aligned, unassociated stars as a binary depends strongly on the pair's position in phase space, so not taking these biases into account may lead to serious problems. At the same time, the nature of Keplerian orbits combined with an assumption about the orbital separations of wide binaries provides us with an expectation about the properties of a distribution of real binaries. 
In this section, we quantify the likelihood that any stellar pair is both a random alignment and a true binary. The combination of these likelihoods with their associated prior probabilities provides the Bayesian posterior probability that any pair is a true binary.

As the details of our model are fairly technical, readers may wish to read the overview in Section~\ref{sec:method_overview} and then skip to the discussion of  contaminants in Section~\ref{sec:contamination}. A test of our method on an existing catalog of wide binaries drawn from the rNLTT is presented in Appendix~\ref{sec:rNLTT}.

\subsection{Overview of the Method}
\label{sec:method_overview}
\begin{figure*}
\begin{center}
\includegraphics[width=0.95\textwidth]{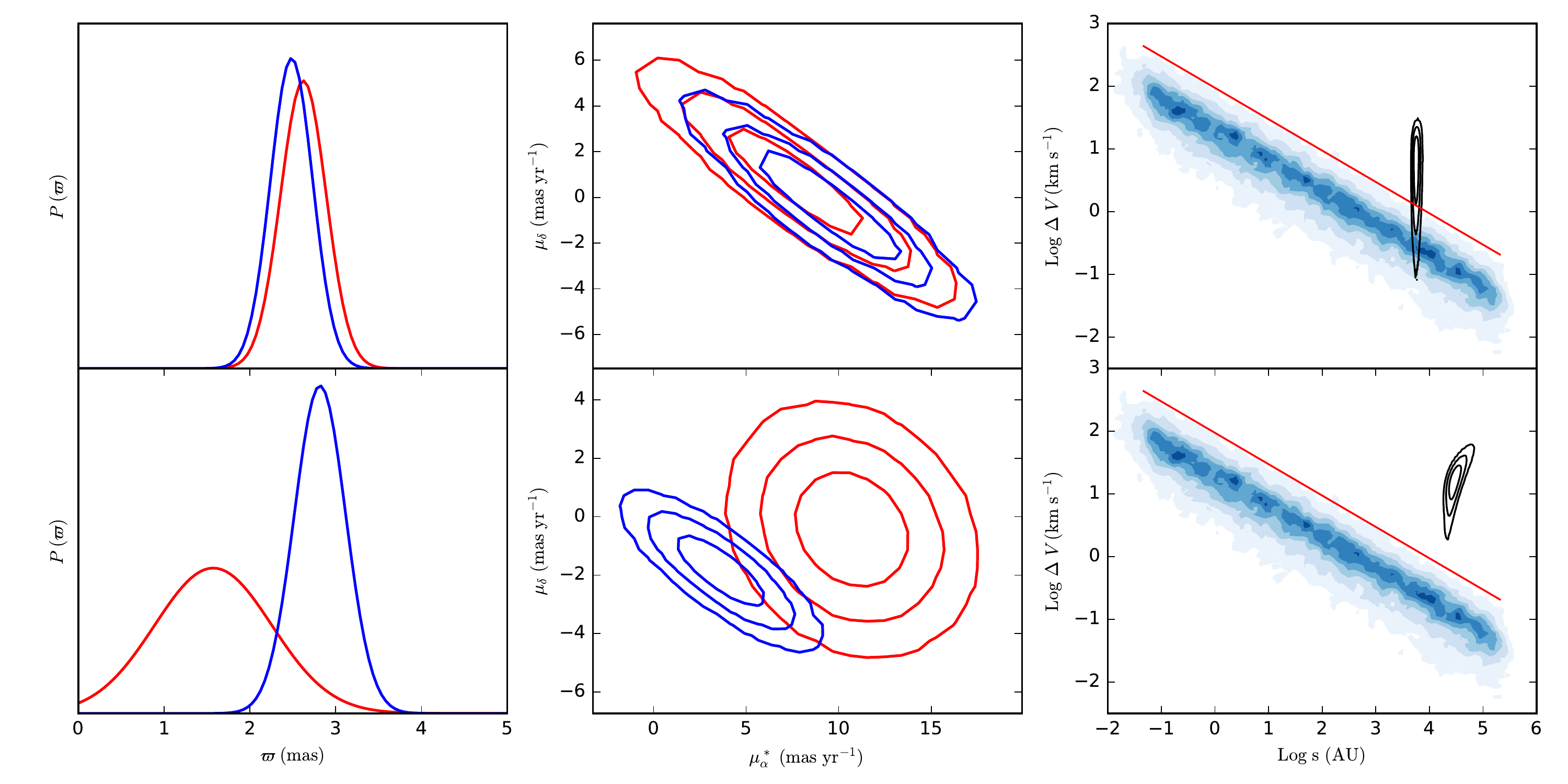}
\caption{The top row are measurements for a pair we identify as a true binary, while the bottom row are for a pair we reject. The left column shows the parallaxes and their uncertainties for each star as red and blue Gaussians. Concentric error ellipses in the middle column compare the proper motions of each star in the pair. In the right column, we compare the position in $\log s-\log \Delta V$ space of two pairs of stars from the TGAS catalog (black contours) to the expectation of a population of binary stars generated by the method described in Section~\ref{sec:challenges} (blue background). The contours, representing 1, 2, and 3$\sigma$ confidence levels, are created by accounting for uncertainties in the parallaxes and proper motions of the stars in each pair. Contours from the pair in the top right panel overlap with the region of parameter space expected for genuine binaries, while contours in the bottom right panel do not. The red lines in the right column are an analytic estimate from Equation \ref{eq:delta_V_limit} for the region above which stellar pairs cannot be bound. }
\label{fig:example_s_dV}
\end{center}
\end{figure*}

Figure \ref{fig:example_s_dV} illustrates part of our analysis for two separate pairs. The top row is for a pair we identify as a true binary, while the bottom row is for a pair that we determine to be a random alignment. The left column compares the astrometric parallaxes (based on their uncertainties) of the two stars in each pair as red and blue Gaussian curves, while the middle column shows the corresponding proper motion uncertainty ellipses for both stars in each pair. Concentric ellipses indicate correlated 1, 2, and 3$\sigma$ regions. The system in the top row clearly has matching astrometry, while in the bottom row the parallaxes and proper motions are only consistent at the $\approx$2$-$3$\sigma$ level.

To generate the plots in the last column of  Figure~\ref{fig:example_s_dV}, we transform each pair's angular separation and each star's parallax and proper motion to $s$ and $\Delta V$. The distribution in these two parameters for both pairs of stars is indicated by the contours, which correspond to  1, 2, and 3$\sigma$ confidence levels, while the blue background in these panels is the expected distribution of binaries simulated in Section~\ref{sec:challenges}. The convolution in $s-\Delta V$ space of the observed with the expected distribution provides the likelihood that the difference in a pair's proper motions is consistent with binary motion. This likelihood, along with a statistic measuring the consistency of the two stars' parallaxes, forms the  probability that a pair's astrometry matches that of a genuine binary. We construct this term mathematically in Section~\ref{sec:binary_likelihood}.

This is one part of the calculation in a Bayesian formalism; we also need to determine the likelihood that any given pair is consistent with being formed from a random alignment of stars. The likelihood function is based on the probability that any two stars can have the observed angular separation, proper motion difference, and parallax; random alignments are more likely to have larger angular separations, larger proper motion differences, and smaller parallaxes. This is one of the crucial aspects for the reliable identification of wide binaries, and we describe how we obtain this term in Section~\ref{sec:random}.

Finally, we need to calculate the prior probabilities of any given pair being a genuine binary and a random alignment of stars. Since the number of stellar binaries in a sample scales with the size of the sample, $N$, while the number of randomly aligned pairs scales with $N(N-1)/2$, any arbitrary pair has a very strong prior of being a random alignment. The exact value of this prior, which depends on the local density of a pair's position and proper motion, needs to be determined individually for every pair. We describe how these priors are calculated in the following section.

\subsection{Constructing the Bayesian Formalism}
\label{sec:bayesian_formalism}

We consider any two stars occupying similar positions in five-dimensional phase space (position: $\alpha$, $\delta$; proper motion: $\mu_{\alpha}$, $\mu_{\delta}$; and parallax: $\varpi$) to be formed from one of two classes: random alignments ($C_1$) or genuine binaries ($C_2$). This probability will, in general, depend on all five parameters. For example, random alignments are more likely to be in dense stellar regions and to involve stars with small proper motions. To account for variations in this probability, we first assume that we are only interested in pairs with similar enough positions and proper motions  that there is no difference in the average stellar density in the vicinity of each star. This is a reasonable assumption for all but the most widely separated stars. This assumption allows us to separate the two sets of astrometric parameters (one for each star) into $\vec{x}_j$, the pair's position in four-dimensional phase space, and $\vec{x}_i$, which contains the scalar difference between the two stars' positions and proper motions as well as each stars' measured parallax:
\begin{eqnarray}
\vec{x}_i &=& \{\theta, \Delta \mu', \varpi'_1, \varpi'_2 \}, \\
\vec{x}_j &=& \{ \alpha, \delta, \mu_{\alpha}, \mu_{\delta} \}.
\end{eqnarray}

We use primes to indicate observed quantities that have some non-negligible uncertainty associated with them: uncertainties in the angular separation, $\theta$, can be ignored while uncertainties in $\mu$ (and hence $\Delta \mu$) and $\varpi$ cannot. 

For computational efficiency, we take the components of $\vec{x}_j$ to be the primary star's position and proper motion. The components of $\vec{x}_i$ will be determined by the difference between the two measured values; each component (save for $\theta$) will have an uncertainty associated with it. For closely separated stars $\theta$ can be determined from the two stellar coordinates:
\begin{equation}
\theta \approx \sqrt{(\alpha_A - \alpha_B)^2 \cos \delta_A \cos \delta_B
			 + (\delta_A - \delta_B)^2}.
\end{equation}

$\Delta \mu$  can be similarly calculated:
\begin{equation}
\Delta \mu \approx \sqrt{(\mu^*_{\alpha, A} - \mu^*_{\alpha, B})^2 
			+ (\mu_{\delta, A} - \mu_{\delta, B})^2}, \label{eq:proper_motion}
\end{equation}
where $\mu^*_{\alpha, i} = \mu_{\alpha, i}\cos \delta_i$.

Using these sets of parameters and the two classes $C_1$ and $C_2$, we can now use Bayes's theorem to construct the generalized probability that any pair  forms a true binary:
\begin{equation}
P(C_2 \given \vec{x}_i, \vec{x}_j) = \frac{P(\vec{x}_i \given C_2, \vec{x}_j) P(C_2 \given \vec{x}_j)}{P(
\vec{x}_i \given \vec{x}_j)}. \label{eq:P_binary_1}
\end{equation}

The first term in the numerator of Equation \ref{eq:P_binary_1} is the likelihood; the second term is the prior probability of a binary with $\vec{x}_j$. The denominator is the evidence, which can be determined by summing the probability that the particular pair can be produced by both a binary ($C_2$) and a random alignment ($C_1$):
\begin{equation}
P(\vec{x}_i \given \vec{x}_j) = \sum_{k=1,2} P(\vec{x}_i \given C_k, \vec{x}_j) P(C_k \given \vec{x}_j).
\end{equation}

We determine the prior probabilities, $P(C_1 \given \vec{x}_j)$ and $P(C_2 \given \vec{x}_j)$, based on our understanding of binaries and random alignments: $P(C_1 \given \vec{x}_j)$ should scale with the square of the stellar density in position and proper motion phase space, $\rho^2(\vec{x}_j)$, while $P(C_2 \given \vec{x}_j)$ should scale with $\rho(\vec{x}_j)$ if we include both resolved and unresolved pairs:
\begin{eqnarray}
P(C_1 \given \vec{x}_j) &=& A_1\ \rho^2(\vec{x}_j) \\
P(C_2 \given \vec{x}_j) &=& A_2\ \rho(\vec{x}_j).
\end{eqnarray}

We determine the coefficients of these equations by integrating the relations over all four dimensions of $\vec{x}_j$, which will provide us with the total number of random alignments in the sample, $N(N-1)/2$, and the total number of binaries in the sample (including unresolved pairs), $f_{\rm bin} N$, where $f_{\rm bin}$ is the binary fraction:
\begin{eqnarray}
\frac{N(N-1)}{2} &=& A_1\ \int \dd \vec{x}_j\ \rho^2(\vec{x}_j) \\
N f_{\rm bin} &=& A_2\ \int \dd \vec{x}_j\ \rho(\vec{x}_j).
\end{eqnarray}

The integral to determine $A_1$ must be calculated numerically for a specific sample, while $A_2$ immediately becomes $f_{\rm bin}$:
\begin{eqnarray}
A_1 &\approx& \frac{1}{2 \int \dd \vec{x}_j\ P(\vec{x}_j)^2} \\
A_2 &=& f_{\rm bin},
\end{eqnarray}
where we have used the substitution $\rho(\vec{x}_j) = N P(\vec{x}_j)$. We calculate the integral over $A_1$ using Monte Carlo random sampling over kernel density estimates (KDEs) of the distribution in position space and in proper motion space. We describe these density estimators in detail in Section \ref{sec:random}.

We can now substitute the priors into Equation \ref{eq:P_binary_1} to obtain our posterior distribution:
\begin{equation}
P(C_2 \given \vec{x}_i, \vec{x}_j) = \frac{P(\vec{x}_i \given C_2, \vec{x}_j) f_{\rm bin} }{P(\vec{x}_i \given C_2, \vec{x}_j) f_{\rm bin}  + P(\vec{x}_i \given C_1, \vec{x}_j) A_1 \rho(\vec{x}_j)  }. \label{eq:P_binary_2}
\end{equation}

We note here that $P(C_2 \given \vec{x}_i, \vec{x}_j)$ is normalized in such a way that the corresponding posterior probability for a particular binary being a random alignment can be expressed as:
\begin{equation}
P(C_1 \given \vec{x}_i, \vec{x}_j) = 1 - P(C_2 \given \vec{x}_i, \vec{x}_j) \label{eq:P_random}
\end{equation}

\subsection{Binary Likelihood: $P(\vec{x}_i \given C_2, \vec{x}_j)$}
\label{sec:binary_likelihood}


We now determine the probability that a true binary could produce the observations, $P(\vec{x}_i \given C_2, \vec{x}_j)$. We begin by accounting for quantities with observational uncertainties. We marginalize over the true proper motion difference $\Delta \mu$, and the true binary parallax, $\varpi$. Both components of a true binary can be approximated as having the same distance. This is because the widest binaries have separations $\sim$1 pc, less than the uncertainty on distances derived from parallax (using {\it Gaia}'s nominal parallax uncertainty of 0.3 mas) for stars at distances beyond $\approx$60 pc. Since, as we will show in Figure \ref{fig:sample_P_posterior}, most binaries are found at significantly smaller separations and larger distances, we can simplify the problem by only marginalizing over the true parallax of the binary rather than each star's parallax independently:
\begin{eqnarray}
P(\vec{x}_i \given C_2, \vec{x}_j) &=& \int \dd \Delta \mu\ \dd \varpi\ P( \Delta \mu, \varpi, \vec{x}_i \given C_2, \vec{x}_j ). \label{eq:P_binary_marginalized_1} \\
&=& \int \dd \Delta \mu\ \dd \varpi\ P(\theta, \Delta \mu, \varpi \given C_2)\ P(\varpi'_1 \given \varpi, \Delta \mu) \nonumber \\
& & \qquad \times\ P(\Delta \mu' \given \varpi, \Delta \mu)\
P(\varpi'_2 \given \varpi, \Delta \mu )  \label{eq:P_binary_marginalized_2}
\end{eqnarray}
We have separated out the observables $\varpi'_1$, $\varpi'_2$, and $\Delta \mu'$ since they are dependent only on their true underlying values and their correlated uncertainties. 

We can factor the first term in the integrand of Equation \ref{eq:P_binary_marginalized_2} so $P(\theta, \Delta \mu, \varpi \given C_2) = P(\theta, \Delta \mu \given C_2, \varpi)\ P(\varpi)$. Our understanding of binaries does not directly yield distributions in $\theta$ and $\Delta \mu$. However, binary theory and observations provide $P(s, \Delta V \given C_2)$, the expected distribution of the projected physical separation, $s$, and the tangential velocity difference, $\Delta V$. These are the physical forms of the angular variables $\theta$ and $\Delta \mu$:  $\theta = s \varpi$ and $\Delta \mu = \Delta V \varpi$. Calculating the first term in Equation \ref{eq:P_binary_marginalized_2} then amounts to performing a Jacobian transformation from $(\theta, \Delta \mu)$ to $(s, \Delta V)$:
\begin{equation}
P(\theta, \Delta \mu, \varpi \given C_2) = P(s, \Delta V \given C_2) \frac{1}{\varpi^2},
\end{equation}
where one factor of $\varpi$ enters in the denominator for each of $s$ and $\Delta V$ due to the Jacobian transformation. Care must be taken to ensure that these two factors of $\varpi$ are in the correct units: one factor should convert from $\Delta V$ to $\Delta \mu$ and the other should convert from $s$ to $\theta$. We can now reduce Equation \ref{eq:P_binary_marginalized_2} to:
\begin{eqnarray}
P(\vec{x}_i \given C_2, \vec{x}_j) &=& \int \dd \Delta \mu\ \dd \varpi\ P(s, \Delta V \given C_2)\ \frac{P(\varpi)}{\varpi^2} P(\varpi'_1\given \varpi, \Delta \mu) \nonumber \\
& & \qquad \times\ P(\Delta \mu' \given \varpi, \Delta \mu)\ 
P(\varpi'_2 \given \varpi, \Delta \mu ) \label{eq:P_binary_marginalized}
\end{eqnarray}

We now focus on the first term in the integrand in Equation \ref{eq:P_binary_marginalized}, $P(s, \Delta V \given C_2)$, which expresses the likelihood that a random binary would produce the observed $s$ and $\Delta V$. In general, this function depends on assumptions made about populations of binary stars. We assume a binary is completely determined by four parameters: the two stellar masses, $M_1$ and $M_2$, the orbital separation, $a$, and the eccentricity. We do not include any binary evolution interactions between the two stars.

We determine the density of true binaries in $s-\Delta V$ space by randomly generating 10$^4$ binaries from two different distributions.\footnote{Our tests indicate that this is a sufficient sample size to balance accuracy with computation time.} The first distribution uses a prior on $a$ that is flat in log space ($P(a)\propto a^{-1}$), following \"{O}pik's Law. We allow $a$ to range between 10 $R_{\odot}$ and 1 pc, so that half of all binaries have $a >$ 10$^2$ AU. The priors for the remaining orbital parameters required to produce distributions in $s$ and $\Delta V$ are described in Section~\ref{sec:challenges}; see also Figure \ref{fig:gaia_binary_limits}.

We also test a model with a steeper power-law prior distribution: $P(a)\propto a^{-1.6}$. This is motivated by the results of \citet[][hereafter CG04]{chaname04} and \citet{lepine07}, who find a steeper power-law-like decline in the distribution of binaries at large separations. As in the log-flat model, $a$ extends to 1 pc; the inner boundary of $\approx$33 AU is chosen so that half of all binaries have $a>$ 10$^2$ AU. Throughout the remainder of the paper, we use both orbital separation models to derive our results. 

The differences between the models are substantial: the log-flat model places relatively more weight at larger separations, increasing the possibility of identifying binaries at wider separations. The cost is an increased rate of contamination at these large separations. We test samples of binaries obtained from both methods to obtain our conclusions on the orbital separation distribution of wide binaries in Sections \ref{sec:orb_sep} and \ref{sec:s_model} to ensure that our results are not dependent upon our adopted model assumptions.

From both populations of binaries, we use a KDE with a top hat kernel to create a normalized probability density function in $s-\Delta V$ space. Evaluating the KDE at a particular $s$ and $\Delta V$ provides $P(s, \Delta V \given C_2)$. 

Circular binaries satisfy the inequalities that $s\leq a$ and $\Delta V^2 \leq G M_{\rm tot} a^{-1}$, where $G$ is the gravitational constant and $M_{\rm tot}$ is the combined binary mass. These can be combined:
\begin{equation}
\Delta V \leq \sqrt{\frac{G M_{\rm tot}}{s}}. 
\label{eq:delta_V_limit}
\end{equation}

Eccentric binaries do not obey this inequality since, for a fraction of the orbit near pericenter, the relative velocity of the two stars can substantially exceed the circular approximation. Furthermore, the mass of an arbitrary binary is unknown, so a fiducial total binary mass must be adopted, and the restriction on $\Delta V$ must be relaxed. 

Nevertheless, we find that when non-circular binaries and a range of masses are taken into account, the observed $\Delta V$ for a given $s$ is typically lower than in the circular case. Guided by our population of randomly generated binaries, we set the maximum binary mass to 10 \Msun\ to account for flexibility in the eccentricity and total binary mass. We show the boundary of this inequality as a red line in Figure~\ref{fig:example_s_dV}. Including this as an additional constraint greatly decreases computation time, as there are fewer calls to the KDE of $P(s, \Delta V \given C_2)$:
\begin{equation}
P(s, \Delta V \given C_2) =
    \begin{cases}
      0 & \Delta V > \Delta V_{\rm max}(s) \\
      KDE(s, \Delta V) & \Delta V \leq \Delta V_{\rm max}(s)
    \end{cases}
\end{equation}
where $\Delta V_{\rm max}(s)$ comes from Equation \ref{eq:delta_V_limit}, where we set $M_{\rm tot}$ to 10 \Msun:
\begin{equation}
\Delta V_{\rm max}(s) = 2.11 \left( \frac{s}{10^3 {\rm AU}} \right)^{-1/2} {\rm km}\ {\rm s}^{-1}
\end{equation}

Returning to Equation \ref{eq:P_binary_marginalized}, the $P(\varpi)$ term is the prior on the parallax measure. This probability distribution is not flat, which is the origin of the Lutz-Kelker bias \citep{lutz73}. However, \citet{smith87} demonstrated that the exact form of this distribution for a magnitude-limited sample differs from the original formulation in \citet{lutz73}. And recently, \citet{astraatmadja16} showed that the distances to stars in the TGAS catalog are reasonably approximated using an exponentially decaying prior distribution with a scale length, $L$, of 1.35 kpc. This prior distribution, when represented in terms of $\varpi$, is:
\begin{equation}
P(\varpi) = \left\{\begin{array}{lc}
 \frac{1}{2 L^3 \varpi^4} \exp\left[ -\frac{1}{L \varpi} \right] & {\rm for}\ 0 < \varpi, \\
 0 & {\rm otherwise}.
 \end{array}
 \right. \label{eq:P_plx}
\end{equation}
We use this form for $P(\varpi)$ throughout this work. An improved model could include the directional dependence of the parallax prior. For instance, the distribution of distances to stars is likely substantially different depending on whether we are looking out of the Galactic Plane or toward the Galactic Center. In practice, our results are largely unaffected by our choice of a parallax prior (see discussion in Section~\ref{sec:assumptions}).

We now return to Equation \ref{eq:P_binary_marginalized}. The integral on the right hand side of the equation is a multidimensional integral of the form: $\int \dd \vec{y}\ P(\vec{z}_j, \given \vec{y}, \vec{z}_i)\ P(\vec{y} \given \vec{z}_i)$, where $\vec{z}_i$ and $\vec{z}_j$ are arbitrary variables. This integral can be approximated using Monte Carlo importance sampling:
\begin{equation}
\int \dd \vec{y}\ P(\vec{z}_j, \given \vec{y}, \vec{z}_i)\ P(\vec{y} \given \vec{z}_i) \approx \frac{1}{N} \sum_k^N P(\vec{z}_j \given \vec{y}_k, \vec{z}_i), \label{eq:importance_sampling}
\end{equation} 
where $\vec{y}_k$ is drawn randomly $N$ times from the joint conditional probability: $\vec{y}_k \sim P(\vec{y} \given \vec{z}_i)$.

Adopting this approximation, Equation \ref{eq:P_binary_marginalized} reduces to:
\begin{equation}
P(\vec{x}_i \given C_2, \vec{x}_j) \approx \frac{1}{N}\sum_k P(s, \Delta V \given C_2)\ \frac{P(\varpi_k)}{\varpi_k^2}
 P(\varpi'_2 \given \varpi_k, \Delta \mu_k ),
 \label{eq:P_binary_approx}
\end{equation}
where $\varpi_k$ and $\Delta \mu_k$ are random samples drawn from their joint conditional probabilities. To obtain appropriately sampled $\varpi_k$ and $\Delta \mu_k$ based on $P(\varpi'_1\given \varpi, \Delta \mu)$ and $P(\Delta \mu' \given \varpi, \Delta \mu)$, we combine these two terms into one multivariate normal distribution ($\mathcal{N}$) and recognize its symmetric nature: 
\begin{eqnarray}
P(\varpi'_1 \given \varpi, \Delta \mu)\ P(\Delta \mu' \given \varpi, \Delta \mu) &=& P(\varpi'_1, \Delta \mu' \given \varpi, \Delta \mu) \nonumber \\
&=& \mathcal{N}(\varpi'_1, \Delta \mu' \given \varpi, \Delta \mu) \nonumber \\ 
&=& \mathcal{N}(\varpi, \Delta \mu \given \varpi'_1, \Delta \mu'), \label{eq:flip_normal}
\end{eqnarray}
where $\Delta \mu'$ is dependent upon the proper motions for both stars.

We can now obtain random samples of $\varpi$ and $\Delta \mu$ by drawing random samples from both stars' observed multivariate normal distributions and their corresponding covariance matrices ($\matr{\Sigma}$):
\begin{eqnarray}
\mu^*_{\alpha, 1,k}, \mu_{\delta, 1,k}, \varpi_{1, k} &\sim& \mathcal{N}\left( \mu^*_{\alpha, 1}, \mu_{\delta, 1},  \varpi_1; \matr{\Sigma_1} \right) \label{eq:star1_random_binary} \\
\mu^*_{\alpha, 2,k}, \mu_{\delta, 2,k}, \varpi_{2, k} &\sim& \mathcal{N}\left( \mu^*_{\alpha, 2}, \mu_{\delta, 2},  \varpi_2; \matr{\Sigma_2} \right). \label{eq:star2_random_binary} 
\end{eqnarray}
Random samples of $\varpi$ are the set of $\varpi_{1, k}$ and random samples of $\Delta \mu$ are the scalar difference between each star's proper motions calculated using the definition in Equation \ref{eq:proper_motion}; we discard random samples of $\varpi_{2, k}$.
Finally, we can calculate $P(\varpi'_2 \given \varpi_k, \Delta \mu_k )$ in Equation \ref{eq:P_binary_approx} by recognizing that it can be inverted into a multivariate normal distribution, as was done in Equation \ref{eq:flip_normal}. Now, we {\it evaluate} the multivariate normal distribution for star 2 at $\varpi'_2$ conditional on the set of {\it already generated} proper motion samples, as well as $\varpi_k$ and $\Delta \mu_k$:
\begin{equation}
P(\varpi'_2 \given \varpi_k, \Delta \mu_k ) = \mathcal{N} \left( \mu^*_{\alpha, 2,k}, \mu_{\delta, 2,k}, \varpi_k; \mu'_{\alpha, 2}, \mu'_{\delta, 2}, \varpi'_2, \matr{\Sigma_2} \right).
\end{equation}

The sum in Equation~\ref{eq:P_binary_approx} provides the likelihood function. Convergence tests indicate that 10$^5$ Monte Carlo random samples of $\varpi$ and $\Delta \mu$ provides sufficient accuracy for our purposes.

\subsection{Random Alignment Likelihood: $P(\vec{x}_i \given C_1, \vec{x}_j)$}
\label{sec:random}

The likelihood that a pair of stars with $\vec{x}_i$ and $\vec{x}_j$ is the product of a random alignment is $P(\vec{x}_i \given C_1, \vec{x}_j)$. We begin by marginalizing over the true individual parallaxes $\varpi_1$ and $\varpi_2$ and the true proper motion difference $\Delta \mu$ to account for observational uncertainties in these quantities:
\begin{equation}
P(\vec{x}_i \given C_1, \vec{x}_j) = \int \dd \varpi_1\ \dd \varpi_2\ \dd \Delta \mu\ P(\varpi_1, \varpi_2, \Delta \mu, \vec{x}_i \given C_1, \vec{x}_j). \label{eq:P_noise_marginalized}
\end{equation}

Now we can substitute for $\vec{x}_i$ and factor out $\varpi'_1$, $\varpi'_2$, and $\Delta \mu'$, as observed quantities are dependent only on their underlying values (and their associated, correlated uncertainties): 
\begin{eqnarray}
P(\vec{x}_i \given C_1, \vec{x}_j) &=& \int  \dd \varpi_1\ \dd \varpi_2\ \dd \Delta \mu\ P(\varpi_1)\ P(\varpi_2) \nonumber \\
	& & \times \
    P(\varpi'_1, \varpi'_2, \Delta \mu' \given \varpi_1, \varpi_2, \Delta \mu) \nonumber \\
	& &  \times\  P(\theta, \Delta \mu \given C_1, \vec{x}_j). \label{eq:RA_substituted}
\end{eqnarray}

Our assumption that $\theta$ and $\Delta \mu$ are independent allows us to split the last term in the integrand. Of course, Galactic structure implies that position and proper motion are not entirely independent. However, accounting for this joint dependence adds a significant degree of complexity and computational expense to the problem. In the test of our method in Appendix \ref{sec:rNLTT}, we recover all the previously identified pairs in the rNLTT catalog, indicating any reduction of the effectiveness of our method due to this approximation is minor. Equation \ref{eq:RA_substituted} becomes:
\begin{eqnarray}
P(\vec{x}_i \given C_1, \vec{x}_j) &=& \int  \dd \varpi_1\ \dd \varpi_2\ \dd \Delta \mu\ P(\varpi_1)\ P(\varpi_2) \nonumber \\
	& & \times \
    P(\varpi'_1, \varpi'_2, \Delta \mu' \given \varpi_1, \varpi_2, \Delta \mu) \nonumber \\
	& &  \times\  
P(\theta \given C_1, \alpha, \delta)\ 
P(\Delta \mu \given C_1, \mu^*_{\alpha}, \mu_{\delta}) \label{eq:RA_factored}
\end{eqnarray}

This integral over $\varpi_1$, $\varpi_2$, and $\Delta \mu$ is of the form shown on the left hand side of Equation \ref{eq:importance_sampling} and can therefore be calculated using Monte Carlo importance sampling:
\begin{eqnarray}
P(\vec{x}_i \given C_1, \vec{x}_j) &\approx& \frac{1}{N} \sum_k 
P(\varpi_{1,k})\ P(\varpi_{2,k})\ P(\theta \given C_1, \alpha, \delta) \nonumber \\
	& & \times \
P(\Delta \mu_k \given C_1, \mu^*_{\alpha}, \mu_{\delta}),
\label{eq:RA_approx}
\end{eqnarray}
where $\varpi_{1,k}$, $\varpi_{2,k}$, and $\Delta \mu_k$ are $N$ samples generated from the probability distribution $P(\varpi'_1, \varpi'_2, \Delta \mu' \given \varpi_1, \varpi_2, \Delta \mu)$. As was done in Equation \ref{eq:flip_normal}, this distribution is equivalent to a multivariate normal distribution, the arguments of which can be flipped:
\begin{equation}
P(\varpi'_1, \varpi'_2, \Delta \mu' \given \varpi_1, \varpi_2, \Delta \mu) = \mathcal{N}(\varpi_1, \varpi_2, \Delta \mu \given \varpi'_1, \varpi'_2, \Delta \mu')
\end{equation}

We generate these random samples in the same way as was done in Section \ref{sec:binary_likelihood}. First, we generate random proper motions and parallaxes for each star using the observed multivariate normal distribution and its covariance matrix using Equations \ref{eq:star1_random_binary} and \ref{eq:star2_random_binary}. $\Delta \mu_k$ is then determined from the scalar difference of randomly selected proper motions of each of the two stars using Equation \ref{eq:proper_motion}.

To determine the $\theta$ term in Equation \ref{eq:RA_approx}, we recognize that as $\theta$ increases, the probability of random alignments increases linearly. This is because the probability of a star having a randomly aligned companion at an angular separation $\theta$ is found from the integrated stellar density around an infinitesimally thin annulus of radius $\theta$. For a population with a linearly increasing density in space (or a uniform density), this probability is proportional to $2 \pi\ \theta$ and the stellar density at the central star's position. So long as changes in the stellar density does not substantially deviate from linearity on spatial scales larger than the typical binary separation, this approximation is appropriate.

Using an analogous argument, the $\Delta \mu$ term in Equation~\ref{eq:RA_factored} should scale linearly with both $\Delta \mu$ and the density in $(\mu_{\rm \alpha^*}, \mu_{\delta})$ space. However, Figure \ref{fig:tycho-2_pos_mu} shows that, contrary to what is seen in position space, the density in proper motion space changes substantially on relatively smaller scales, particularly at smaller $\mu$ where the distribution is strongly peaked. We tested our prescription by finding the typical $\Delta \mu$ at which the number of stars surrounding a point significantly deviates from our linear approximation; while most wide binaries have $\Delta \mu \sim 1-3$ mas yr$^{-1}$, our approximation is typically valid for $\Delta \mu < 5$ mas yr$^{-1}$. Nevertheless, our locally linear approximation may become inaccurate in specific cases. We leave a higher-order extension of this probability for future work.

The resulting conditional probabilities for $\theta$ and $\Delta \mu$ can now be expressed as:
\begin{eqnarray}
P(\theta \given C_1, \alpha, \delta) &=& \frac{2 \pi}{N} \rho(\alpha, \delta)\ \theta. \nonumber \\
 &=& 2 \pi\ KDE(\alpha, \delta)\ \theta \label{eq:P_theta} \\
P(\Delta \mu \given C_1, \mu^*_{\alpha}, \mu_{\delta}) &=& \frac{2 \pi}{N} \rho_{\mu}(\mu^*_{\alpha}, \mu_{\delta})\ \Delta \mu, \nonumber \\
  &=& 2 \pi\ KDE(\mu^*_{\alpha}, \mu_{\delta})\ \Delta \mu \label{eq:P_mu}
\end{eqnarray}
where $\rho(\alpha, \delta)$ and  $\rho_{\mu}(\mu^*_{\alpha}, \mu_{\delta})$ are the positional and proper motion-dependent densities, respectively, and $N$ is the number of stars in the catalog and serves as a normalization constant.

To calculate these densities, we generate two separate KDEs with top hat kernels for stars in the TGAS catalog: one in $(\alpha \cos \delta, \delta)$ space, and one in $(\mu^*_{\alpha}, \mu_{\delta})$ space. The KDEs act as smoothing functions that interpolate over the distributions in the two panels in Figure \ref{fig:tycho-2_pos_mu} and produce a normalized probability density function in position and $\mu$ space.

We use the KDE implementation in the {\tt Python} package {\tt scikit-learn}. This algorithm offers a substantial efficiency gain as it is based on a tree method, which has a computational time scaling with $O(N \log M)$, where $N$ is the number of function calls and $M$ is the number of elements in the tree, rather than the $O(NM)$ scaling of standard KDEs. We further optimize our algorithm by building the KDE tree based on a randomly chosen subset of 10$^5$ TGAS stars rather than for the full TGAS catalog. The size of this subset and the KDE bandwidth are optimized to minimize computation time while maintaining structure in the position and proper motion distributions.

The two parallax terms of form $P(\varpi)$ in Equation~\ref{eq:RA_approx} represent the prior probability on parallax in the TGAS catalog. We use the exponentially decaying prior provided in Equation \ref{eq:P_plx}.

Using our Monte Carlo sampling from Equations \ref{eq:star1_random_binary} and \ref{eq:star2_random_binary}, and substituting for the functions in Equations \ref{eq:P_plx}, \ref{eq:P_theta} and \ref{eq:P_mu}, we can now use the sum in Equation \ref{eq:RA_approx} to calculate $P(\vec{x}_i \given C_1, \vec{x}_j)$, the likelihood that any particular pair of stars is the product of a random alignment. Our convergence tests indicate that 10$^5$ random samples provide sufficient accuracy for our purposes here.

\begin{table*}
\begin{center}
\caption{The astrometric parameters of the 12 open clusters from which we remove candidate pairs. RA, Dec, and $R$ provide the position and angular radius around which we identify stars associated with these clusters. $\mu_{\rm RA}$, $\mu_{\rm Dec}$, and $\Delta \mu$ define the corresponding proper motion constraint, and $\varpi$ and $\Delta \varpi$ define the parallax of each cluster. Note that these are ``by-eye'' approximations to {\it Gaia} astrometry rather than precisely measured characteristics. \label{tab:open_clusters}}
\begin{tabular}{lccr@{.}lr@{.}lr@{.}lr@{.}lr@{.}lc} 
\toprule
Cluster & RA & Dec &  \multicolumn{2}{c}{$R$} &  \multicolumn{2}{c}{$\mu_{\rm RA}$} &  \multicolumn{2}{c}{$\mu_{\rm Dec}$} & \multicolumn{2}{c}{$\Delta \mu$} &  \multicolumn{2}{c}{$\varpi$} & $\Delta \varpi$ \\
  & [deg.] & [deg.] &  \multicolumn{2}{c}{[deg.]} &  \multicolumn{2}{c}{[mas yr$^{-1}$]} &  \multicolumn{2}{c}{[mas yr$^{-1}$]} & \multicolumn{2}{c}{[mas yr$^{-1}$]} &  \multicolumn{2}{c}{[mas]} & [mas] \\
\midrule
Pleiades & 03:46:00 & $+$24:06:00 & 8&6 & 20&10 & $-$45&39 & 10&0 & 7&4 & 2.0 \\ 
Coma Ber & 12:24:00 & $+$26:00:00 & 7&5 & $-$11&75 & $-$8&69 & 10&0 & 11&53 & 3.0 \\
Hyades & 04:27:00 & $+$15:52:12 & 18&54 & 110&0 & $-$30&0 & 30&0 & 21&3 & 5.0 \\ 
Praesepe & 08:40:00 & $+$19:42:00 & 4&5 & $-$35&81 & $-$12&85 & 10&0 & 5&49 & 2.0 \\ 
$\alpha$ Per & 03:30:00 & $+$49:00:00 & 8&0 & 22&73 & $-$26&51 & 5&0 & 5&8 & 2.0 \\ 
IC 2391 & 08:40:00 & $-$53:06:00 & 8&0 & $-$24&69 & 22&96 & 10&0 & 6&90 & 2.0 \\ 
IC 2602 & 10:40:48 & $-$64:24:00 & 2&9 & $-$17&02 & 11&15 & 5&0 & 6&73 & 2.0 \\ 
Blanco I & 00:04:24 & $-$30:06:00 & 2&5 & 20&11 & 2&43 & 5&0 & 4&83 & 2.0 \\ 
NGC 2451 & 07:41:12 & $-$38:30:00 & 2&0 & $-$21&41 & 15&61 & 5&0 & 5&45 & 2.0 \\ 
NGC 6475 & 17:53:36 & $-$34:48:00 & 3&5 & 2&06 & $-$4&98 & 10&0 & 3&70 & 2.0 \\ 
NGC 7092 & 21:31:36 & $+$48:24:00 & 3&5 & $-$8&02 & $-$20&36 & 5&0 & 3&30 & 2.0 \\ 
NGC 2516 & 07:57:36 & $-$60:42:00 & 3&5 & $-$4&17 & 11&91 & 7&0 & 2&92 & 2.0 \\
\bottomrule
\end{tabular}
\end{center}
\end{table*}

\section{Characterizing Contamination}
\label{sec:contamination}

We address the problem of separating genuine wide binaries from randomly aligned stars in the TGAS catalog. Our approach is to first characterize the population of contaminating objects, which may be co-moving stars in kinematic structures such as open clusters, or, more likely, randomly aligned, unassociated stars.

\subsection{Identifying and Removing Open Clusters}
\label{sec:open clusters}

In addition to applying our statistical method constructed in Section \ref{sec:method}, we remove any pairs consistent with being members of 12 open clusters: the Pleiades, Coma Ber, Hyades, Praesepe, $\alpha$ Per, IC 2391, IC 2602, Blanco I, NGC 2451, NGC 6475, NGC 7092, and NGC 2516. We rely largely on \citet{vanLeeuwen09} to define the astrometric positions of these clusters, although we update the values based on {\it Gaia}'s improved astrometry (we adjusted the distance to the Pleiades, for instance). We remove stars that simultaneously satisfy a position, proper motion, and parallax constraint: they are within the cluster radius $R$ and have a proper motion and parallax within the cluster proper motion width $\Delta \mu$ and a parallax width of $\Delta \varpi$. Table~\ref{tab:open_clusters} shows the parameters used to identify stars in these 12 open clusters. Note that these are not precisely measured cluster characteristics; they are ``by-eye'' approximations to the {\it Gaia} astrometry of several known open clusters. 

Additional clusters and moving groups are likely to be found in the TGAS catalog; indeed, \citet{oh2016} identify dozens of such co-moving groups composed of three or more elements in the TGAS catalog. Such structures typically span many square degrees on the sky, and are found with projected separations $>$1 pc. Our prior on the projected separation of wide binaries excludes co-moving stars with these separations from being identified by our algorithm, and we do not expect co-moving groups to be a strong source of contamination for our sample. We discuss this possibility further in Section \ref{sec:1pc_binaries}.

\subsection{Constructing a Sample of Random Alignments}
\label{sec:random_alignments}

\begin{figure*}
\begin{center}
\includegraphics[width=0.95\textwidth]{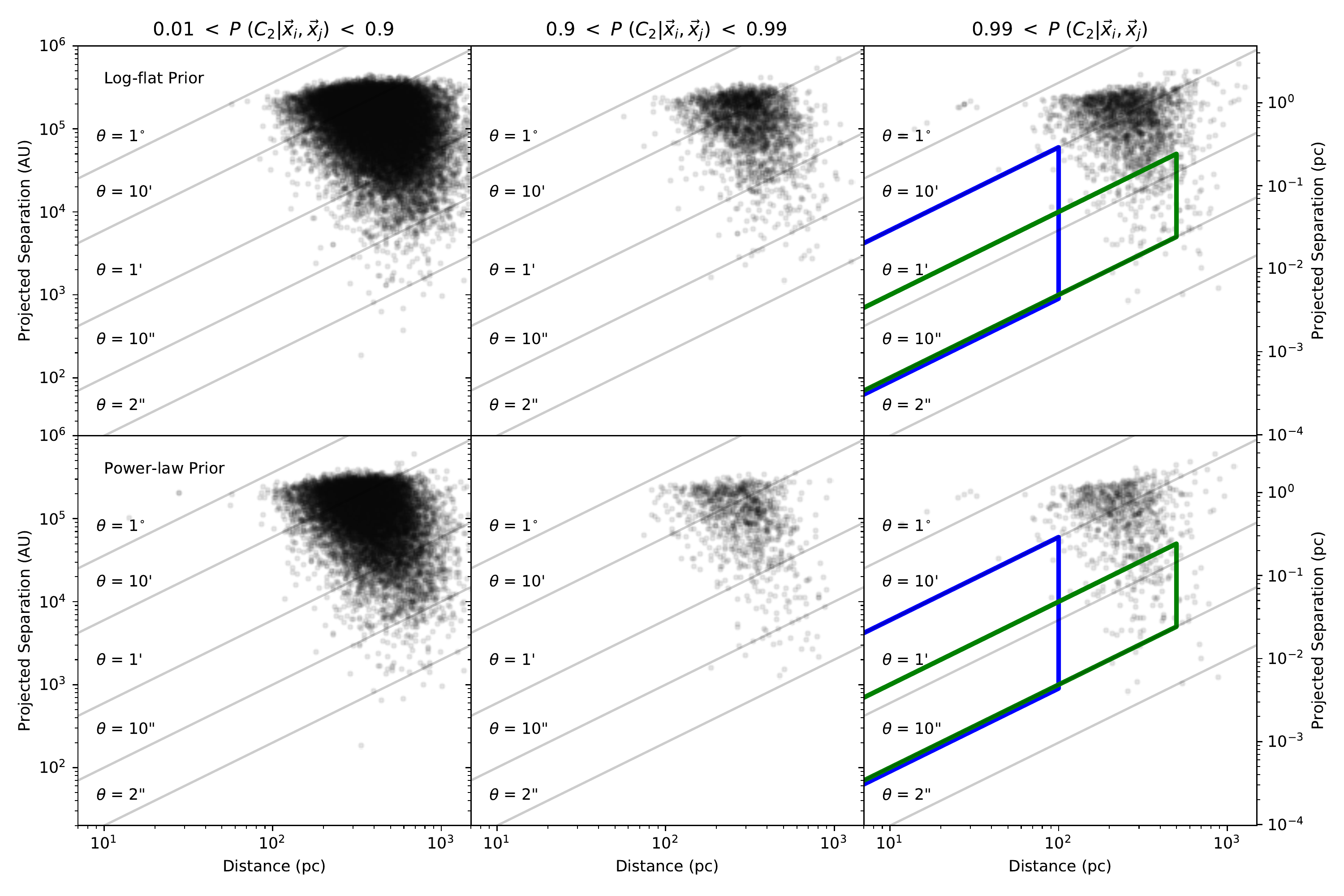}
\caption{ Distances and projected separation $s$ of random alignments generated by matching the stars in the TGAS catalog to a shifted version of the same catalog. Lines of constant $\theta$, which are the actual observable, are overlaid. The pairs in these panels define the locus of randomly aligned stars. Relatively few random alignments fall within the blue and green boxes. We note that these panels show that even individual pairs with angular separations as small as a 1\asec\ may be comprised of a random alignment. Confirming the binarity of a particular stellar pair requires precise RV measurements.}
\label{fig:TGAS_dist_s_theta_false}
\end{center}
\end{figure*}

To characterize the contamination from random alignments, we generate a sample of false pairs by taking the entire TGAS catalog and displacing it in both position {\it and} proper motion space, by $+$1\degree in $\delta$ and $+$3 mas yr$^{-1}$ in both $\mu^*_{\alpha}$ and $\mu_{\delta}$. We then match each star in this shifted catalog to the original TGAS catalog; the resulting matches are a catalog in which {\it every} pair is a random alignment. The size of this shift is chosen to be large enough that genuine pairs are not accidentally matched in our false catalog, but small enough that the local densities in position and proper motion space around each star are not changed substantially. This technique is an adaptation of that employed by \citet{lepine07}.

Figure \ref{fig:TGAS_dist_s_theta_false} shows the distances and projected separations for our sample of random alignments; the top row is the model with a log-flat prior on $a$, while the bottom row is for a power-law prior. We use the average parallax, weighted by the parallax uncertainties, to calculate the distance to each pair. The three columns show an increasing posterior probability from left to right; that random alignments populate the panel with a posterior probability of 99\% indicates that contamination exists even at the highest of posterior probabilities. Importantly, random alignments may exist with separations down to the angular resolution of the instrument. The green and blue regions in Figure \ref{fig:TGAS_dist_s_theta_false} define regions in distance and $\theta$ that are relatively free of contamination; we discuss these regions further in Section \ref{sec:identifying_sample}.

From Equation \ref{eq:P_random} the posterior probability of a particular pair being a random alignment is equal to $1 - P(C_2 \given \vec{x}_i, \vec{x}_j)$. Our model assigns a probability $<$1\% of being a random alignment to all those pairs in the rightmost panels of Figure \ref{fig:TGAS_dist_s_theta_false}, yet by construction none of them are binaries. Future improvements to our model may help address these imperfections. For instance, our model does not currently account for geometric effects from the non-Euclidean equatorial coordinates which can lead to slight problems in the proper motion matching for the most widely separated pairs, particularly at large declinations.\footnote{To illustrate this effect, imagine two stars with widely separated right ascensions at high declinations moving directly North. The two stars will have non-parallel proper motions.} We assume that these two distributions are independent, but the proper motion distribution may vary with position. To account for this, a four-dimensional KDE could be generated based on both position and proper motion.

\section{Constructing and Validating the Catalog}
\label{sec:gaia}

\begin{figure}
\begin{center}
\includegraphics[width=1.03\columnwidth]{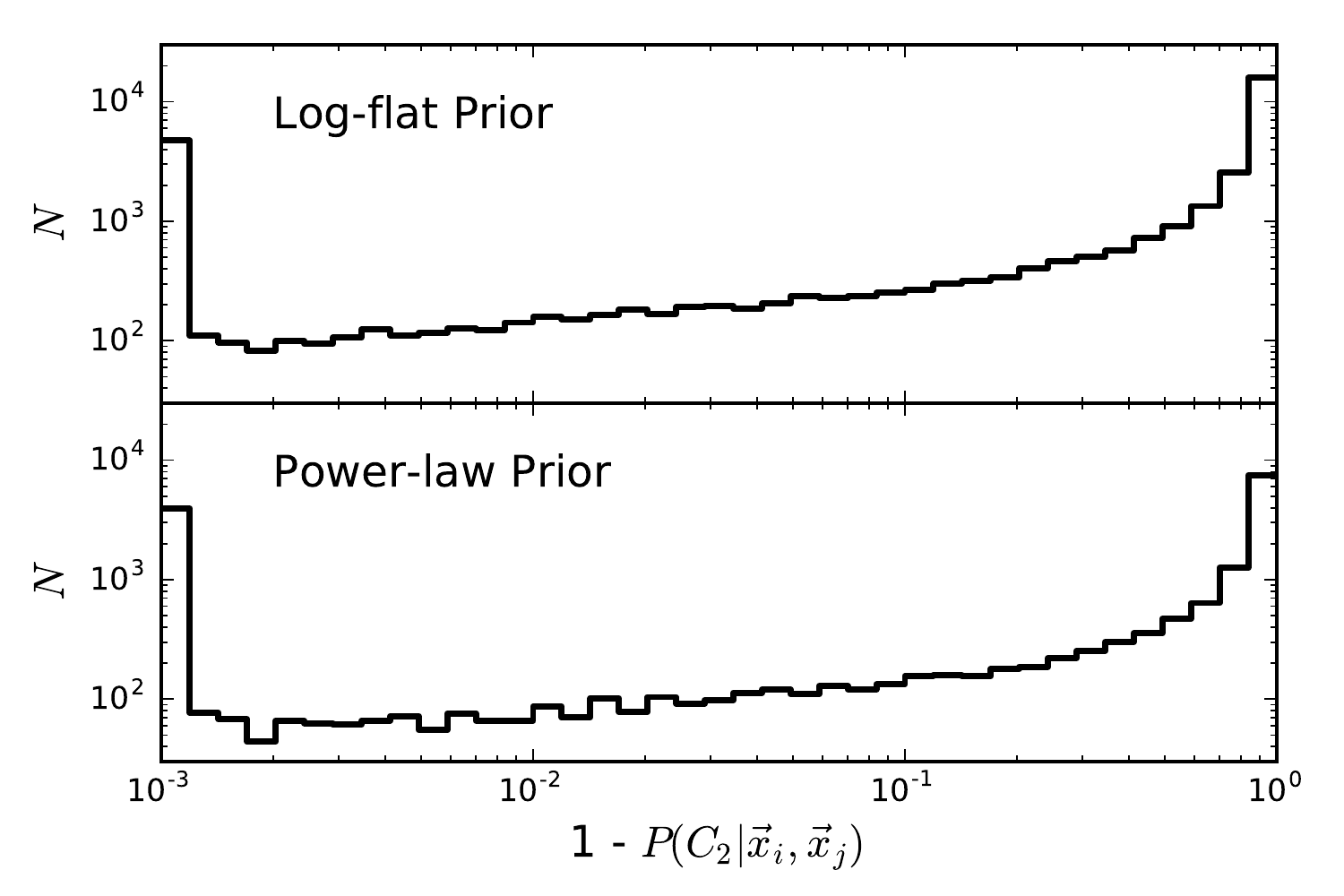}
\caption{ The probability distribution of the pairs with posterior probabilities above 1\% identified using the two different priors for the orbital separation distribution of binaries. Most pairs are identified with a posterior probability either near unity or near zero. Because of this trough in the distribution between the genuine pairs on the left and random alignments toward the right, our resulting sample of wide binaries is largely insensitive to our choice of probability cut-off. }
\label{fig:sample_P_posterior}
\end{center}
\end{figure}

\subsection{Identifying Wide Binaries}
\label{sec:identifying_sample}
 
We now apply our method to the 2,057,050 stars in TGAS. Matching the entire catalog takes $\approx$300 CPU hours. Our algorithm is designed to be embarrassingly parallel, and future searches with larger catalogs can take advantage of large, high-performance computing clusters.

\begin{figure*}
\begin{center}
\includegraphics[width=0.95\textwidth]{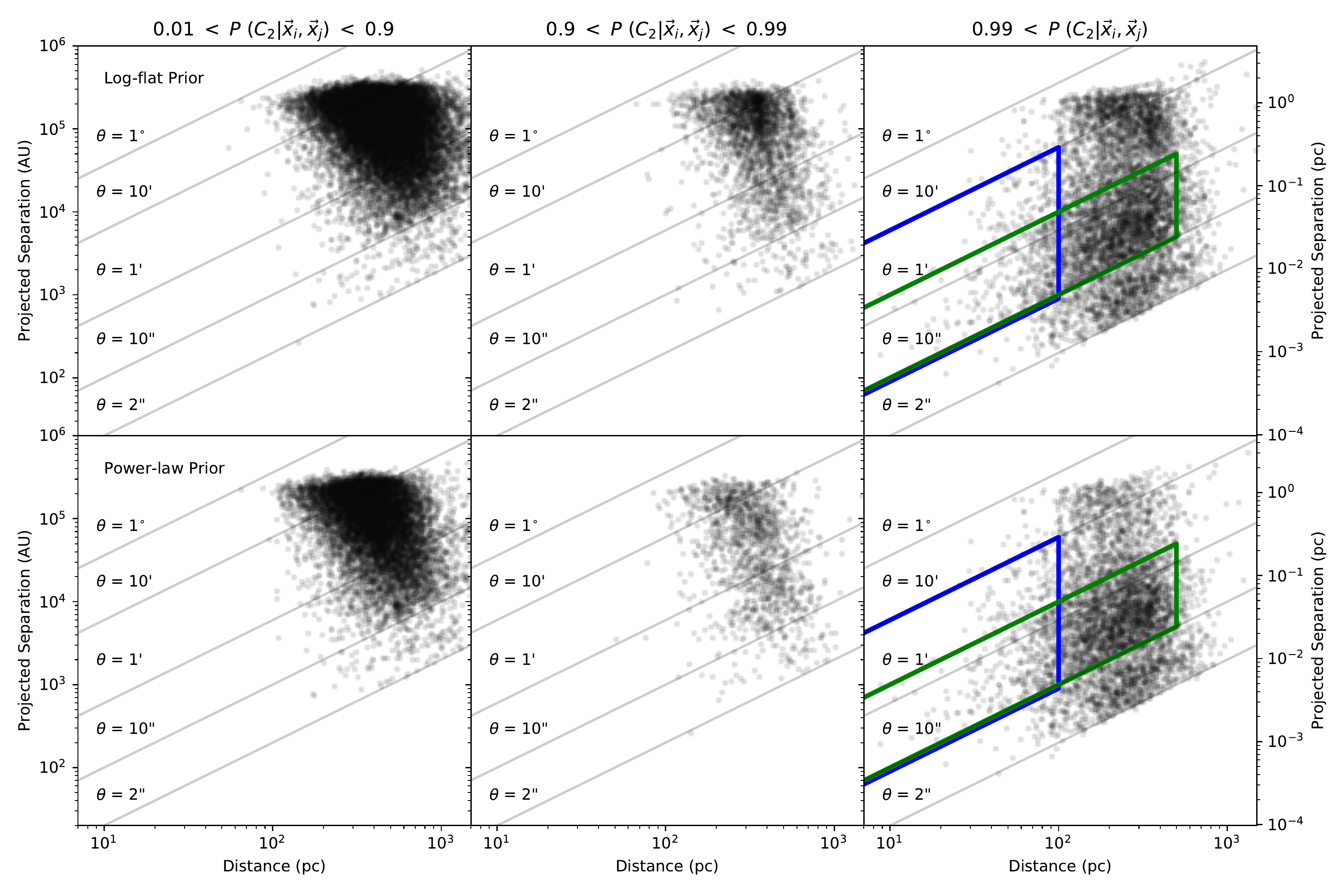}
\caption{As in Figure~\ref{fig:TGAS_dist_s_theta_false}, but for candidate pairs in our two samples identified using a log-flat and a power-law prior on the orbital separation. The left panels shows the distribution of pairs with a posterior probability from 1\% to 90\%. The majority of these pairs are at large distances and separations, indicating that they are almost all random alignments. The middle two panels, showing pairs matched with probabilities between 90\% and 99\% are still dominated by random alignments; however, there may be some genuine pairs in this bin at distances beyond a few 100 pc and separations within 1\amin. The right panels show systems with a posterior probability above 99\% and have a clear population of genuine pairs with smaller separations than the locus of random alignments. In this bin, we define two separate regions with minimal contamination by random alignments: Region 1 with $10\asec < \theta < 100\asec$ and $D < 500$ pc (green lines) and Region 2 with $10\asec < \theta < 10\amin$ and $D < 100$ pc (blue lines).}
\label{fig:TGAS_dist_s_theta}
\end{center}
\end{figure*}

After removing several hundred binaries consistent with being in open clusters, our algorithm identifies 33,169 and 17,895 stellar pairs with posterior probabilities above 1\% using the log-flat and power-law priors, respectively, on the orbital separation distribution, as described in Section \ref{sec:binary_likelihood}.\footnote{Although our prior probabilities are calculated for pairs with $a\leq1$ pc, non-zero eccentricities may allow us to detect some binaries with projected separations $>$1 pc.}

Figure~\ref{fig:sample_P_posterior} shows the distribution of posterior probabilities for the two sets of pairs. The posterior probabilities tend to be either very close to unity or very close to zero. This indicates that although we must choose a critical posterior probability above which to define a candidate wide binary, the resulting catalog is relatively insensitive to that choice.

\subsubsection{Distances, Angular Separations, and Physical Separations}
\label{sec:dist_s_theta}

Figure \ref{fig:TGAS_dist_s_theta} shows the distance, angular separation, and projected physical separation of the candidate wide binaries identified by our method. Distances are determined for each pair from the average of the two stars' parallaxes, weighted by their uncertainties. As in Figure~\ref{fig:TGAS_dist_s_theta_false}, the different columns correspond to samples with different posterior probabilities. 

Figure~\ref{fig:TGAS_dist_s_theta_false} and
Figure~\ref{fig:TGAS_dist_s_theta} show remarkable similarities in their left columns, where binaries have posterior probabilities between 1\% and 90\%: in all four panels there is a locus of pairs whose density increases at both larger $s$ and larger distances. This suggests that the vast majority of pairs in this column in Figure~\ref{fig:TGAS_dist_s_theta} are due to random alignments.

Restricting the samples of false and real binaries to 90\% $< P(C_2 \given \vec{x}_i, \vec{x}_j) <$ 99\% substantially reduces the total number of pairs, but their distributions in the middle panels of Figures \ref{fig:TGAS_dist_s_theta_false} and \ref{fig:TGAS_dist_s_theta} are still very similar. This similarity between the false and real binaries implies that vast majority of these are also random alignments of stars, although there may an excess of pairs at smaller $s$ in the real data, suggesting that some of these may be genuine binaries.

The locus of random alignments in the right column in Figure \ref{fig:TGAS_dist_s_theta_false} at larger $s$ and larger distances also exists in our sample of candidate binaries in the right column of Figure \ref{fig:TGAS_dist_s_theta}. However, the distribution of our candidate binaries with $P(C_2 \given \vec{x}_i, \vec{x}_j)>99$\% includes an additional locus of points with $s \lesssim 4\times10^4$ AU. Since these data do not exist in our sample of random alignments in Figure \ref{fig:TGAS_dist_s_theta_false}, they are the genuine binaries in our sample.
Furthermore, comparing Figures \ref{fig:TGAS_dist_s_theta_false} and \ref{fig:TGAS_dist_s_theta} shows that few genuine binaries are lost by requiring that pairs have $P(C_2 \given \vec{x}_i, \vec{x}_j)>99$\%.

\begin{figure}
\begin{center}
\includegraphics[width=0.95\columnwidth]{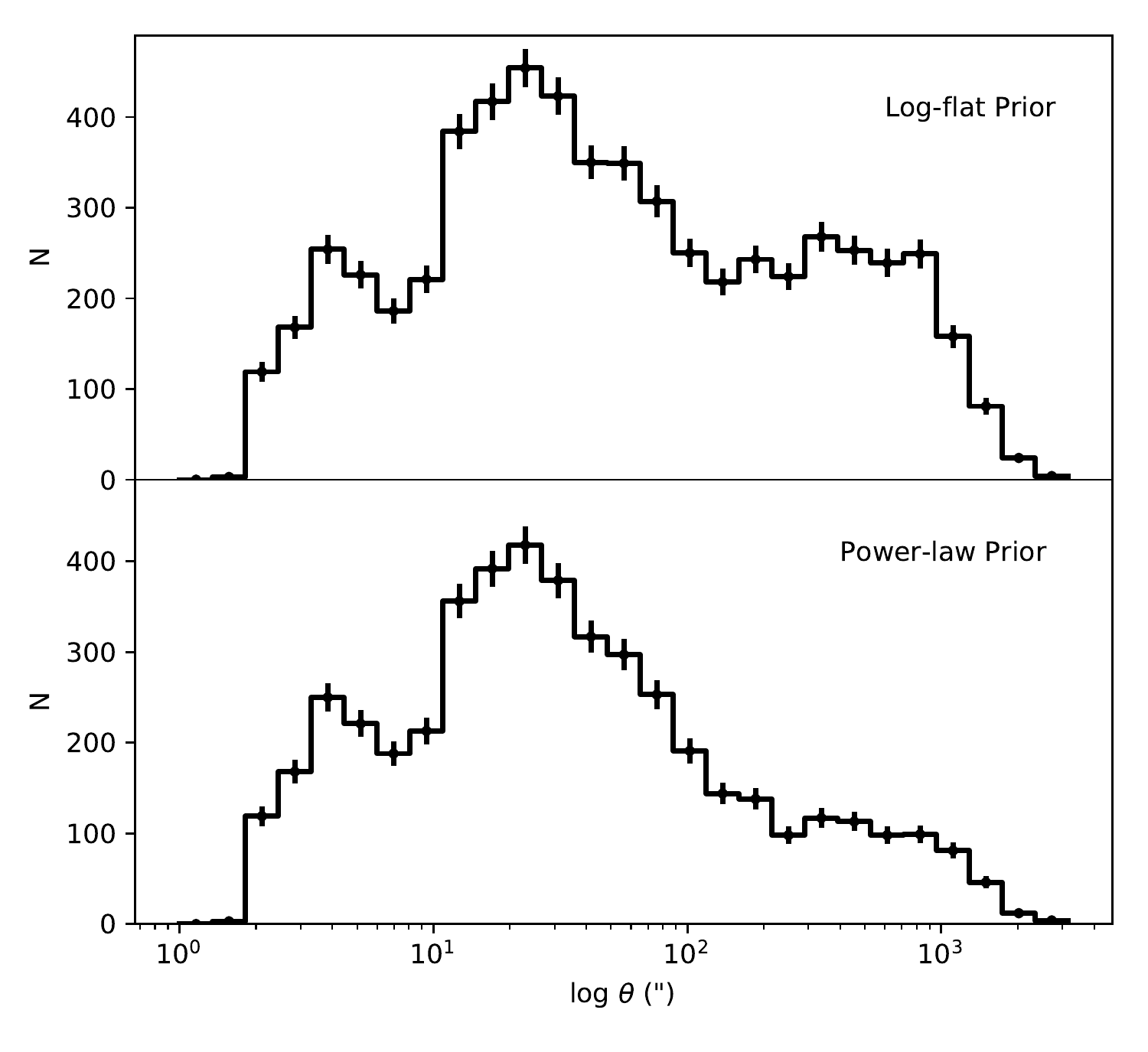}
\caption{ The angular separation distribution of all pairs with a posterior probability above 99\% for a log-flat (top) and a power-law prior (bottom). The steep drop off in the distribution at $\theta<10\asec$ is due to the limits of the Tycho-2 survey, while the excess of systems with $\theta > 200\asec$ is due to contamination by random alignments. Except for pairs with the largest $\theta$, the distributions are largely identical in both structure and number: the locus of points with $s < 4\times 10^4$ AU, where the majority of the genuine wide binaries exist, is $\approx$90\% identical between the two samples. }
\label{fig:TGAS_theta_distribution}
\end{center}
\end{figure}

Grey lines in Figure \ref{fig:TGAS_dist_s_theta} show lines of constant angular separation. The TGAS catalog has a minimum angular separation cutoff at $\approx$2\asec, but there appears to be a substantial decrease in the population at separations $<$10\asec. This is clearly seen by the sharp transition at 10\asec\ in Figure \ref{fig:TGAS_theta_distribution}, which shows the angular separation distribution of our samples from the two methods. Since this decrease scales with $\theta$ rather than $s$, it cannot be an intrinsic property of wide binaries. Instead, this transition is most likely due to the input characteristics of the Tycho-2 catalog, rather than our search method. Future {\it Gaia} data releases will be sensitive to double stars within a fraction of 1\asec\ \citep{debruijne15}.

\begin{figure*}
\begin{center}
\includegraphics[width=0.75\textwidth]{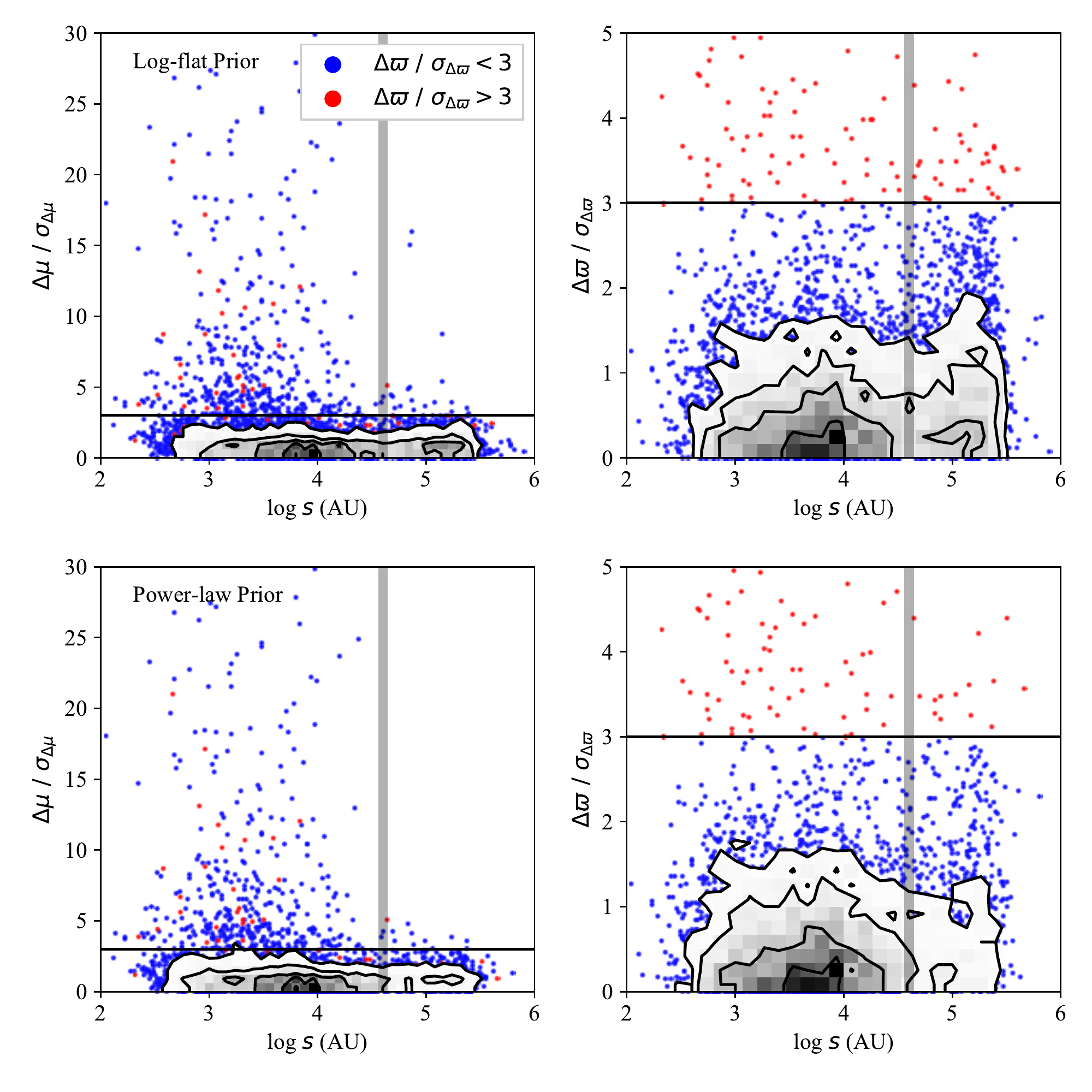}
\caption{ The distribution of proper motion differences (left panels) and parallax differences (right panels) as a function of projected separations for binaries with $P(C_2 \given \vec{x}_i, \vec{x}_j)>99$\% are shown as both points and greyscale contours. Blue (red) points in both panels are those pairs in which the stars have consistent (inconsistent) parallaxes at the 3 $\sigma$ level. The top panels show the sample produced using the log-flat prior while the bottom panels show the sample produced using the power-law prior. The $y$ axes are normalized to their uncertainties; $\sigma_{\Delta \mu}$ is calculated from the quadrature sum of the component uncertainties on $\mu^*_{\alpha}$ and $\mu_{\delta}$ for both components, and $\sigma_{\Delta \varpi}$ is calculated from the quadrature sum of the individual (Gaussian) component uncertainties on $\varpi$. The left panels show that many genuine binaries exist with proper motions that differ by greater than 3 $\sigma$ (horizontal line), justifying our claim in Section \ref{sec:challenges} and predicted by Figure \ref{fig:gaia_binary_limits}: {\it Gaia}'s astrometry is precise enough that to robustly identify wide binaries, the differential proper motion due to orbital motion must be accounted for. The vertical gray line is at 4$\times10^4$ AU, the projected separation below which our sample is almost entirely composed of genuine pairs, as we demonstrate in Section \ref{sec:contamination_rate}. }
\label{fig:TGAS_mu_theta_compare}
\end{center}
\end{figure*}

\begin{figure*}
\includegraphics[width=0.75\textwidth]{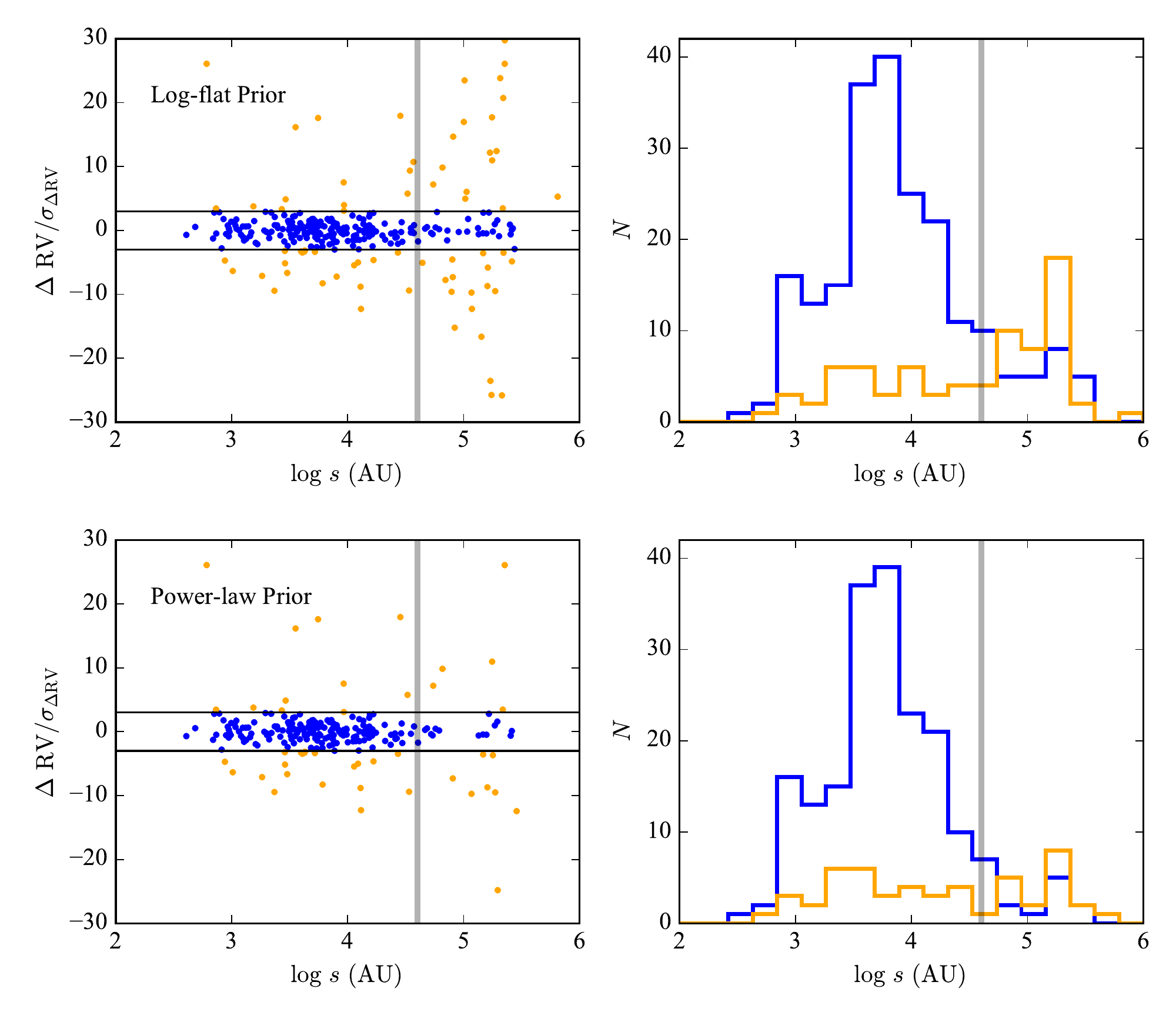}
\caption{$\Delta$RV for the candidate TGAS wide binaries with $P(C_2 \given \vec{x}_i, \vec{x}_j)>$ 99\% and RAVE matches, normalized to its uncertainty, as a function of projected separation for both the log-flat prior (top panels) and power-law prior (bottom panels) models. Candidate pairs with $\Delta$RV less (more) than 3$\sigma$ are shown as blue (orange) points. The right two panels, which compare the projected separation distributions, indicate that random alignments begin to dominate the overall distribution of systems at $s \approx 4\times 10^4$ AU for both models. Restricting for systems with projected separations smaller than this limit (shown by the gray vertical lines) substantially reduces the contamination. }
\label{fig:contamination}
\end{figure*}

An additional test of our wide binary sample can be made by comparing the scalar proper motion difference of the two stars in each of the pairs. The left panels of Figure \ref{fig:TGAS_mu_theta_compare} show this difference, normalized by the proper motion uncertainties, estimated using the quadrature sum of the uncertainties on $\mu^*_{\alpha}$ and $\mu_{\delta}$ for each star, as a function of the projected separation. The set of binaries in our sample with $P(C_2 \given \vec{x}_i, \vec{x}_j)>99$\% are shown both as points and the greyscale contours. The bulk of the pairs are consistent within 3 $\sigma$ (black horizontal line). However, it is immediately apparent that a substantial fraction of binaries (particularly those at smaller projected separations, where contamination by random alignments is negligible; see Figure \ref{fig:contamination}) have component proper motions that deviate from each other at a greater than 3 $\sigma$ significance. This confirms our claim in Section \ref{sec:challenges} and predicted by Figure \ref{fig:gaia_binary_limits}: {\it Gaia's astrometric precision is fine enough that it can detect the orbital motion of a wide binary with a separation of 10$^4$ AU, or equivalently an orbital period of 10$^6$ yr}. Binaries with the most significant proper motion differences are typically {\it Hipparcos} stars with proper motion uncertainties smaller than $\approx0.25$ mas yr$^{-1}$. A wide binary sample produced from a method that does not account for this differential proper motion will fail to identify a large number of wide binaries. 

In the right panels of Figure \ref{fig:TGAS_mu_theta_compare}, we show the corresponding plot for the parallax difference between the two components of the wide binaries, normalized by the uncertainty in the parallax differences, as a function of the projected separations. As expected, most binaries have parallaxes that differ by less than 1 $\sigma$. There is evidence that the parallax difference increases somewhat at separations larger than 4$\times10^4$ AU (vertical gray line), where the sample is expected to be dominated by random alignments. Stellar pairs with parallaxes inconsistent at the greater than 3 $\sigma$ level (but are nevertheless identified by our model with $P(C_2 \given \vec{x}_i, \vec{x}_j)>99$\%) are indicated as red points in both the left and right panels.

\subsubsection{RVs as a Test and Calibration of Our Method}
\label{sec:RV}

While our method for identifying candidate wide binaries relies on five of the six dimensions of phase space, the components of genuine wide binaries should also stand out observationally in the remaining phase space dimension. They should share  essentially the same RVs, whereas randomly aligned pairs should have a random distribution of RVs. Differences in the RVs of the components of a genuine wide binary should be of order $\Delta V$, which, as Figure~\ref{fig:gaia_binary_limits} shows, is $<<$km s$^{-1}$ for all but the closest wide binaries in this sample.

The Fifth Data Release (DR5) of the RAdial Velocity Experiment (RAVE; \citealt{kunder16}) provides RVs for $>$250,000 Southern stars in TGAS.  The typical accuracy of these RVs is better than 2~km~s$^{-1}$. We download the crossmatch between TGAS and RAVE included in RAVE DR5 and eliminate duplicate entries by retaining the observation with highest spectroscopic S/N. This produces a catalog of 210,368 stars, which we search for RV information for the random alignments constructed in Section~\ref{sec:random_alignments}.

\begin{figure*}
\includegraphics[width=0.55\textwidth, angle=-90]{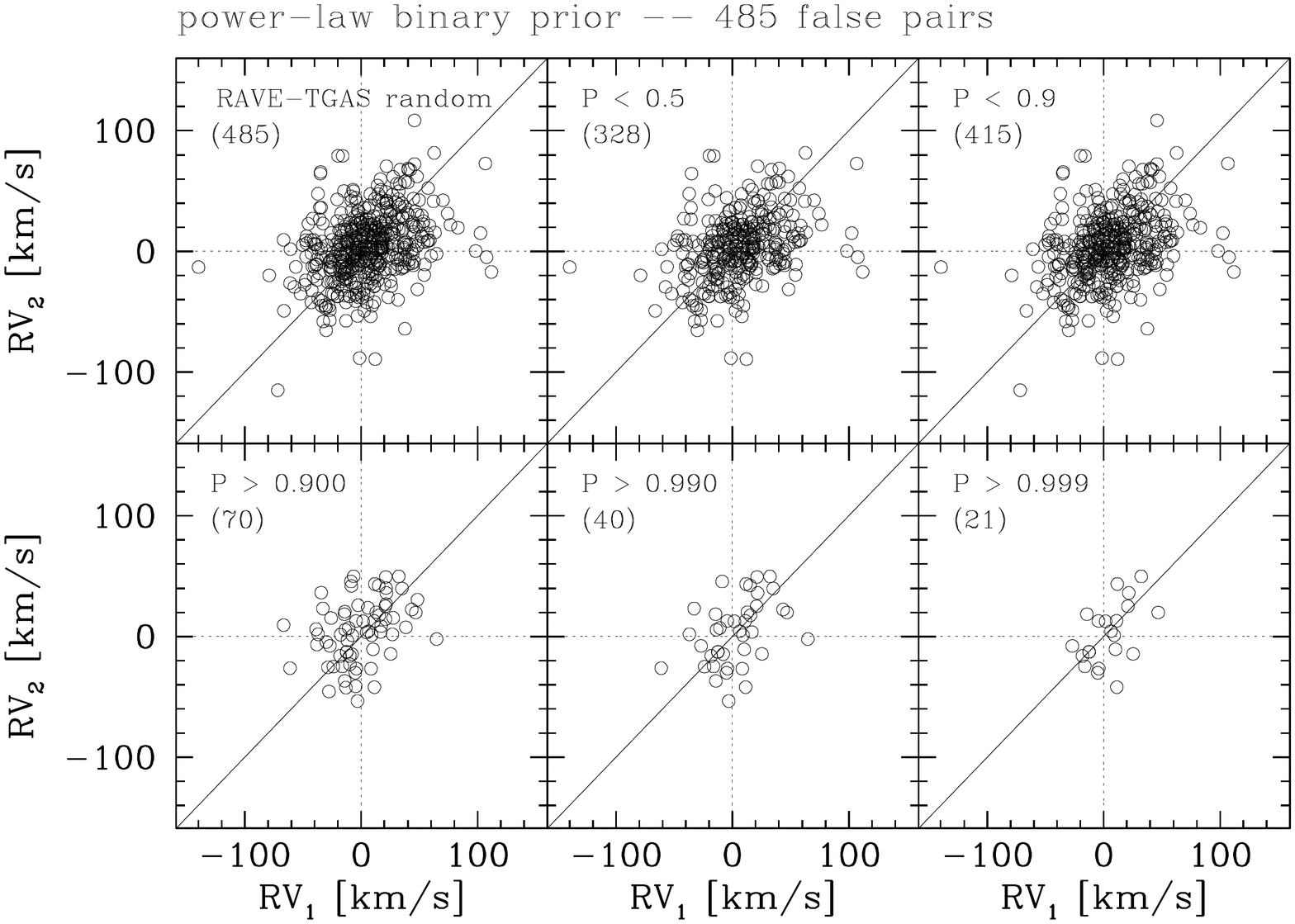}
\caption{Comparison of the RVs of the components of randomly aligned pairs obtained by matching the TGAS catalog to a shifted version of itself. Typical RV uncertainties are smaller than the plot symbols, indicating these pairs are indeed comprised of randomly aligned stars. Note that there is some correlation between the two RVs of our sample of randomly aligned stars, possibly due to our procedure for generating these false pairs by selecting  members of the same kinematic component of the Galaxy. }
\label{fig:deltaRV_false}
\end{figure*}

In Figure~\ref{fig:deltaRV_false}, we compare the RVs of each star in the sample of random alignments generated using the power-law prior for the orbital separation distribution. The typical RV uncertainties are smaller than the plot symbols; clearly these pairs have inconsistent RVs. Figure~\ref{fig:deltaRV_false} shows some non-zero covariance in the distribution of samples.\footnote{One possible reason for this may be due to our matching algorithm selecting members of the same kinematic component of the Galaxy, e.g. thin or thick disk.} Also, we note that matching RVs is a necessary but not sufficient indicator of any particular pair being a genuine binary: Figure~\ref{fig:deltaRV_false} shows that $\approx$20\% of randomly aligned stellar pairs with $P(C_2 \given \vec{x}_i, \vec{x}_j)>$ 99\% happen to have RVs consistent at the 3$\sigma$ level. 

\begin{figure*}
\includegraphics[width=0.55\textwidth, angle=-90]{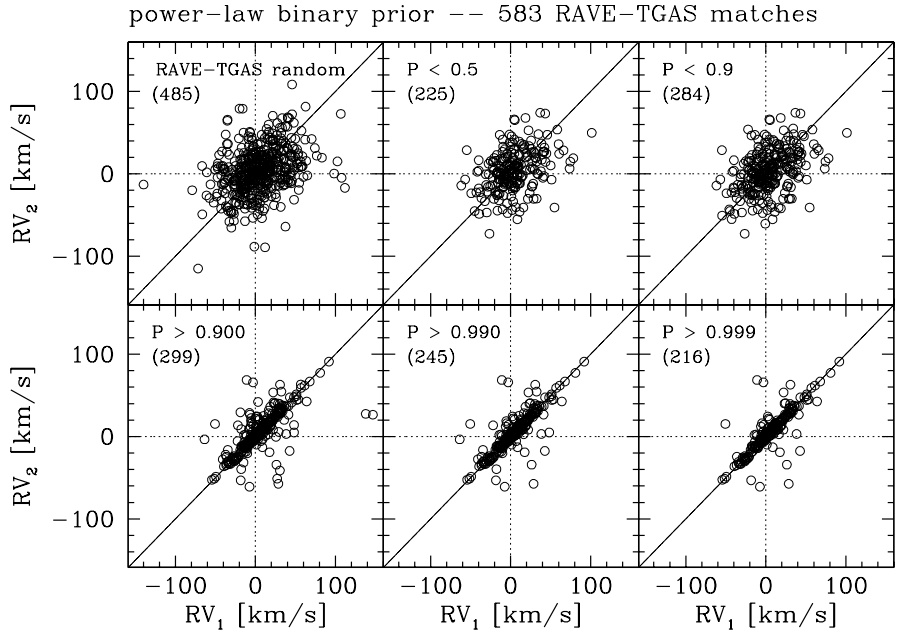}
\caption{RAVE DR5 RVs for the components of wide binary candidates in TGAS identified using the power-law prior for the distribution of binary orbital separations.  The top left panel shows the distribution of completely random pairs in RAVE DR5 and is the same as top left panel in Figure~\ref{fig:deltaRV_false}.  Note that for $P(C_2 \given \vec{x}_i, \vec{x}_j)<$ 90\%, the RV distribution is consistent with that of the random pairings, but for $P(C_2 \given \vec{x}_i, \vec{x}_j)>$ 90\% it falls along the line for equal RVs, becoming progressively narrower as the posterior probability increases.}
\label{fig:RVmatch_powerlaw}
\end{figure*}

\begin{figure*}
\includegraphics[width=0.55\textwidth, angle=-90]{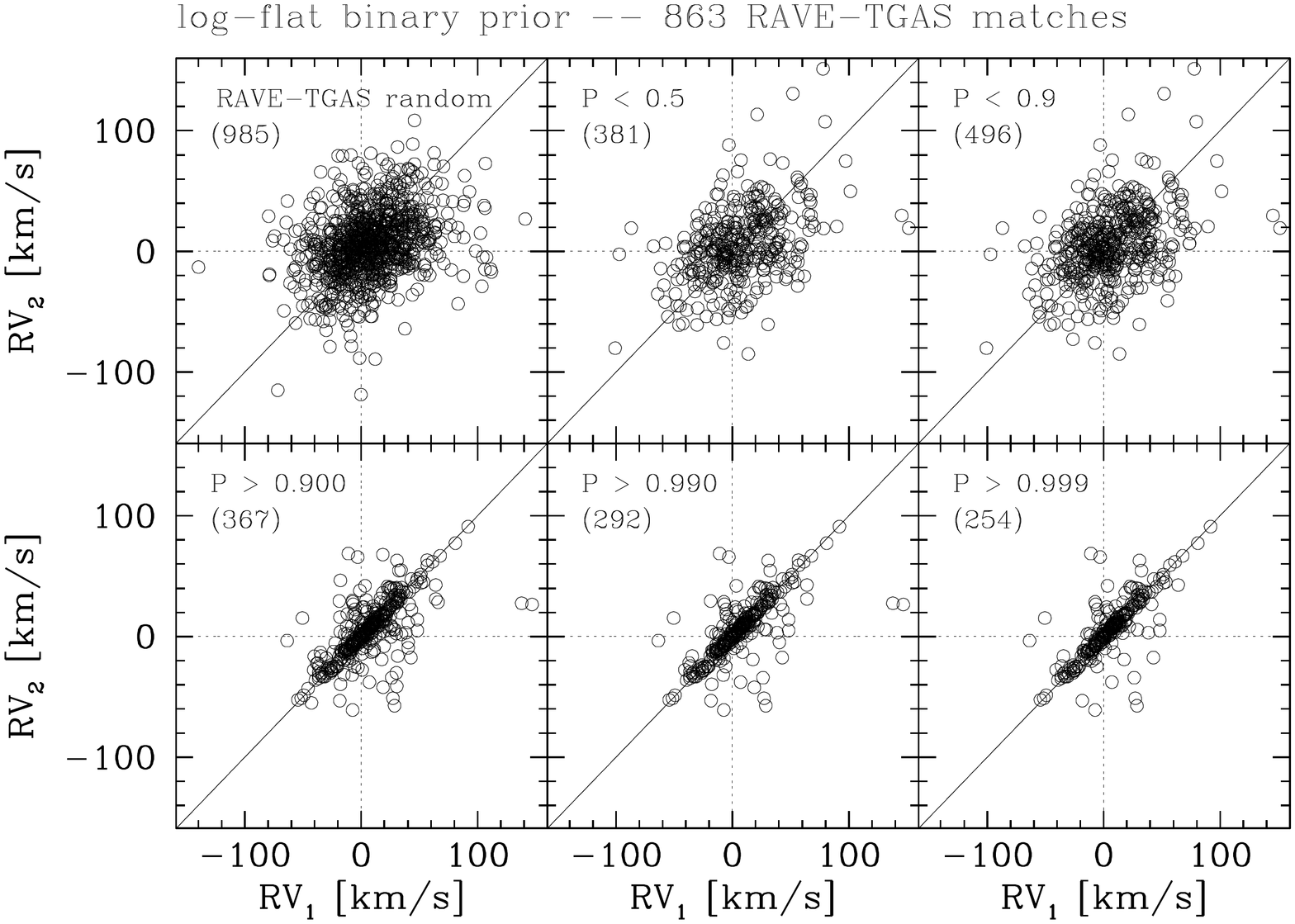}
\caption{Same as Figure \ref{fig:RVmatch_powerlaw} but for the catalog of wide binary candidates identified using the log-flat prior.}
\label{fig:RVmatch_flat}
\end{figure*}

In Figure \ref{fig:RVmatch_powerlaw} we show, plotted against each other, the RAVE RVs for 583 pairs from our 17,895 candidate wide binaries obtained using the power-law prior. We divide the posterior probability $P(C_2 \given \vec{x}_i, \vec{x}_j)$ into five bins and give the number of pairs in each bin. The total number of pairs with RAVE RVs for both components is more than an order of magnitude larger than that identified by \citet{oh2016} for systems with separations $<$1 pc.

\begin{figure*}
\includegraphics[width=0.55\textwidth, angle=-90]{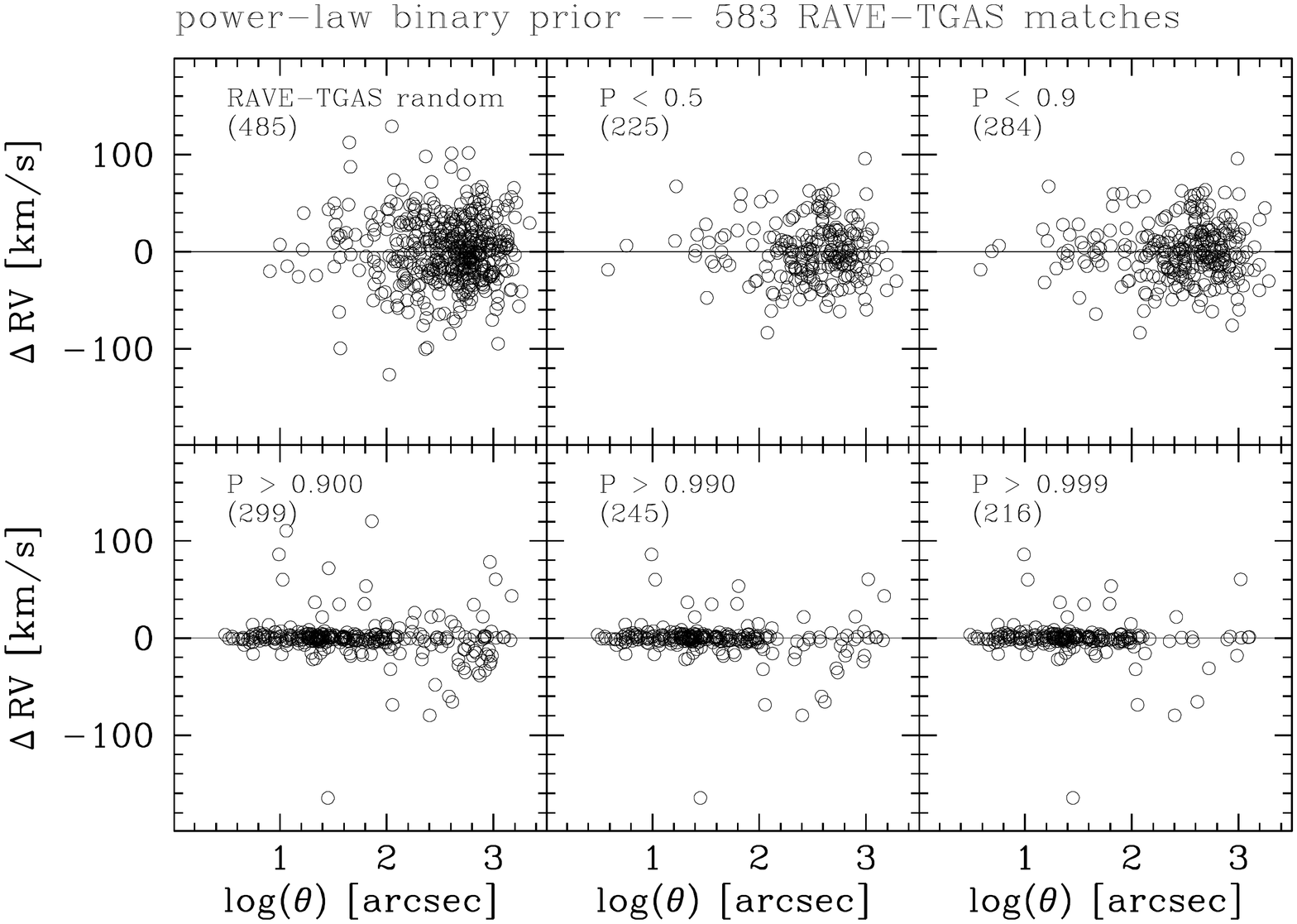}
\caption{$\Delta$RV as a function of angular separation for the components of candidate pairs obtained using the power-law prior for the binary separation distribution.  The distributions for candidate pairs with low probability ($P(C_2 \given \vec{x}_i, \vec{x}_j)<$ 90\%) are systematically wider than those for candidates with high probabilities.}
\label{fig:deltaRV_theta}
\end{figure*}

The sequence of panels in Figure \ref{fig:RVmatch_powerlaw} shows that our candidate pairs progressively align along the line of equal RV as $P(C_2 \given \vec{x}_i, \vec{x}_j)$ increases, unambiguously telling us that genuine wide binaries are being identified with increasingly higher confidence, and thus validating our method. The top middle and top right panels, with $P(C_2 \given \vec{x}_i, \vec{x}_j)<$ 50\% and 90\%, respectively, still look very similar to the plot showing random alignments. When restricting to $P(C_2 \given \vec{x}_i, \vec{x}_j)>$ 90\%, however, the behavior is very different, and a large majority of the 299 candidate pairs in this range have the RVs of the two components falling very close to the one-to-one line. 

However, from our comparison between the distributions in Figures \ref{fig:deltaRV_false} and \ref{fig:RVmatch_powerlaw}, candidate pairs with posterior probabilities between 90\% and 99\% may still include some random alignments, in agreement with our conclusions when comparing Figures \ref{fig:TGAS_dist_s_theta_false} and \ref{fig:TGAS_dist_s_theta}. Therefore, to the limits of the available data, selecting candidates with $P(C_2 \given \vec{x}_i, \vec{x}_j) > 99$\% is likely to result in the purest sample of genuine wide binaries.

Figure~\ref{fig:RVmatch_flat} is the same as Figure~\ref{fig:RVmatch_powerlaw}, but for 863 pairs from our 33,169 candidate wide binaries obtained using the log-flat prior. Again, this sample is an order of magnitude larger than that found by \citet{oh2016} for the same range of separations. In general terms, the same conclusions obtained from Figure~\ref{fig:RVmatch_powerlaw} can be obtained from Figure~\ref{fig:RVmatch_flat}, the main difference being the different relative numbers of candidate pairs falling in each probability range.

In the top left panel of Figure \ref{fig:deltaRV_theta}, we show the difference in RVs of the two stellar components from our sample of false binaries with RAVE RVs described in Section \ref{sec:RV}. These show that the population of random alignments increases at large $\theta$, and is typically of order tens of km s$^{-1}$.\footnote{Given that use of the log-flat prior produces more contamination at low probabilities, we focus on the catalog obtained with the power-law prior for the distribution of orbital separations. Our conclusions are not altered if inspecting instead the catalog obtained using the log-flat prior.}

Here again, the panels of Figure \ref{fig:deltaRV_theta} corresponding to the catalog from the power-law prior show the progressive narrowing of the distribution of pairs around the equal velocity line (in this case, $\Delta$\,RV = 0) as the posterior probability increases, with the overall behavior being markedly different on either side of $P(C_2 \given \vec{x}_i, \vec{x}_j) = 90$\% probability, i.e., a broad distribution of pairs for $P(C_2 \given \vec{x}_i, \vec{x}_j)<$ 90\%, and a significantly narrower one for $P(C_2 \given \vec{x}_i, \vec{x}_j)>$ 90\%.

An additional difference is evident when looking at the distributions as a function of angular separation: candidate pairs are more broadly distributed at low probabilities than they are at the highest ones. Visual inspection of the panels in the upper and lower rows in Figure~\ref{fig:deltaRV_theta} reveals this without ambiguity. This is to be expected if the majority of pairs with $P(C_2 \given \vec{x}_i, \vec{x}_j)<$ 90\% are contamination from random alignments: the probability that two unassociated stars appear to be co-moving increases with the square of the angular separation. In this way, Figure \ref{fig:deltaRV_theta} also argues that the catalogs are dominated by random alignments below 90\% probability, and by genuine binaries at higher probabilities.

Finally, while the assumption that the components of a true wide binary share the same RV holds for most systems, there is one exception: the possibility that one of the components of a genuine wide binary is a (closer) binary system itself. Such hierarchical triples or multiples are not uncommon among wide binaries \citep{makarov08, tokovinin14a,tokovinin14b,tokovinin15,halbwachs17}, and the orbit of the inner subsystem can make the RVs of the components differ from each other when the comparison is based on a single epoch. We further discuss this possibility, as well as other caveats to using RAVE RVs as a calibrator for our method, in Section \ref{sec:contamination_rate}. For now, we note only that these exceptions are a minority of cases, and matching the RVs of the components of a wide binary provides a satisfactory calibration method for our catalogs.

\subsubsection{The Critical Posterior Probability}

From all the evidence just discussed, it seems adequate to require $P(C_2 \given \vec{x}_i, \vec{x}_j)>$ 90\% for a candidate pair to be reliably regarded as a genuine wide binary. However, our goal is to identify as clean a sample of genuine wide binaries as  possible, with no strong selection effects, and to characterize them as a population. Our comparison of sample distances, angular separations, and projected separations for the random alignments  and our candidate binaries in Figures \ref{fig:TGAS_dist_s_theta_false} and \ref{fig:TGAS_dist_s_theta} demonstrate that most of the pairs with 90\% $<P(C_2 \given \vec{x}_i, \vec{x}_j)<$ 99\% are random alignments. With this in mind, therefore, we require $P(C_2 \given \vec{x}_i, \vec{x}_j)>$ 99\% for the wide binary samples we report and analyze in this paper.

Even at $P(C_2 \given \vec{x}_i, \vec{x}_j)>$ 99\%, Figures \ref{fig:RVmatch_powerlaw}, \ref{fig:RVmatch_flat},  and \ref{fig:deltaRV_theta} show a small number of pairs that deviate significantly from the one-to-one and $\Delta$RV = 0 lines. As pointed out earlier, these pairs could represent contamination from chance alignments or be hierarchical triple systems. These should be followed up, ideally with multi-epoch RV measurements, but this is beyond the scope of the present work.

\subsection{Estimating the Contamination Rate}
\label{sec:contamination_rate}

The subset of pairs with RAVE RVs provides us with an estimate for the rate of contamination due to random alignments. In the left two panels of Figure \ref{fig:contamination}, we show the distribution of projected separations for candidates with $P(C_2 \given \vec{x}_i, \vec{x}_j)>99$\% as a function of their RV difference. This difference is normalized by the quadrature sum of the individual RV uncertainties, $\sigma_{\Delta {\rm RV}}^2 = \sigma_{\rm RV,1}^2 + \sigma_{\rm RV,2}^2$, so that deviations from zero correspond to confidence levels. Horizontal lines separate those pairs with consistent RVs at the 3$\sigma$ confidence level (blue points) from those with inconsistent RVs (orange points). The ratio of the number of candidate pairs with discrepant RVs to the overall number of candidates with RAVE matches provides an estimate of contamination. This method assumes that the subset of our candidate binaries with RAVE  matches is a representative sample.

The right two panels of Figure \ref{fig:contamination} show the distribution of projected separations for pairs with consistent RVs (blue) and those with discrepant RVs (orange). Contamination exists at all separations, but the samples are clearly dominated by pairs with consistent RVs out to $s\approx 4\times 10^4$~AU. At larger separations, contamination begins to dominate both samples. Returning to the left two panels of Figure \ref{fig:contamination} and selecting only those systems with $s < 4\times 10^4$~AU (delineated by the vertical gray lines), we estimate the contamination fraction at 17\% for the log-flat prior (37/224) 16\% for the power-law prior (33/212) models.

We can obtain an additional estimate of the contamination fraction using our sample of random alignments. The sample shown in Figure \ref{fig:TGAS_dist_s_theta_false} produces approximately the same number of overall candidate pairs with $P(C_2 \given \vec{x}_i, \vec{x}_j)>1$\% as our search for real wide binaries (both samples are dominated by random alignments with $P(C_2 \given \vec{x}_i, \vec{x}_j)<<99$\%). Restricting for those systems with $P(C_2 \given \vec{x}_i, \vec{x}_j)>99$\% and $s < 4\times 10^4$ AU, we estimate the contamination fraction using this method to be 6\% (245/4360) and 4\% (156/4024) for the log-flat and power-law prior models, respectively.

How can the two methods for estimating the contamination fraction disagree so strongly for the fraction of subset of systems with $s < 4\times 10^4$ AU? Four factors can potentially make the RV-based estimate inaccurate. First, any star in a wide binary that includes an unresolved binary can make a genuine system appear as contamination using RVs. \citet{makarov08} suggest that the fraction of common proper motion pairs that are actually higher order systems may be as high as 25\%. However, if only 10\% of the systems in our sample were triples with an unresolved binary component, our contamination estimate would drop to $\approx$5\%, in agreement with the estimate from our random alignments.

Second, a consistent RV does not guarantee binarity; even systems with RVs consistent at the 3$\sigma$ level may, in fact, be random alignments with similar, but slightly different RVs. Third, if the sample of candidate pairs for which we found matches in RAVE is not a representative TGAS sample, any estimate based on it may be inaccurate. Finally, if the RAVE RV uncertainties are somewhat underestimated, the contamination fraction would be overestimated.

Our contamination fraction estimate using randomly aligned pairs avoids these particular uncertainties. The procedure generating these false pairs is identical to that producing our catalog of candidate binaries. Furthermore, we find that restricting for those candidate pairs with $P(C_2 \given \vec{x}_i, \vec{x}_j)>99$\%, both the false catalog and our catalog of wide binaries produces a comparable number of systems with $s > 10^5$ AU, supporting the idea that the false pair catalog provides a reasonable estimate of contamination for pairs with $s < 4\times 10^4$ AU. With these these caveats in mind, we estimate the contamination fraction to be $\approx$5\%.

Regions 1 and 2 in Figure~\ref{fig:TGAS_dist_s_theta}, outlined by the green and blue boxes, are designed to produce subsets of $P(C_2 \given \vec{x}_i, \vec{x}_j)>99$\% pairs within defined ranges of distances $D$ and angular separations that have very low contamination. Limited to $10\asec < \theta < 100$\asec\ and $D < 500$~pc, Region~1 takes advantage of the large sample of genuine binaries at smaller separations. Region~2 is confined to a distance of 100 pc, while extending to 10\amin\ to test the distribution of binaries at relatively wider separations. We estimate the contamination by counting the number of random alignments falling in these regions. For the log-flat prior model, our random alignment test generates 135 false pairs in Region 1 and four false pairs in Region 2, whereas our search identifies 2704 and 461 candidate pairs in these regions. This corresponds to contamination rates of $\approx$5\% and 1\%, respectively. For the power-law prior, the same test results in nearly identical contaminations rates of $\approx$4\% and 1\% for the two regions (89/2498 4/458). It is from these minimally contaminated subsets that we derive our constraints on the orbital separation distribution of wide binaries.

\restylefloat{table}
\begin{sidewaystable}
\centering
\vspace{-3in}  
\caption{The first 10 wide binaries from our catalog. We provide the {\it Gaia} source IDs, astrometry, posterior probabilities for both the log-flat prior and power-law prior models, and angular separation. This table is available online in its entirety. \label{tab:catalog}}
\resizebox{\textwidth}{!}{
\begin{tabular}{ccccccccccccccccc}
\toprule
\footnotesize
Source ID & Tycho-2 ID & $\alpha_1$ & $\delta_1$ & $\mu^*_{\alpha, 1}$ & $\mu_{\delta, 1}$ & $\varpi_1$ & Source ID & Tycho-2 ID & $\alpha_2$ & $\delta_2$ & $\mu^*_{\alpha, 2}$ & $\mu_{\delta, 2}$ & $\varpi_2$ & $P(C_2 \given \vec{x}_i, \vec{x}_j)$ & $P(C_2 \given \vec{x}_i, \vec{x}_j)$ & $\theta$ \\
 &  &  &  & [mas yr$^{-1}$] & [mas yr$^{-1}$] & [mas] &  &  &  &  & [mas yr$^{-1}$] & [mas yr$^{-1}$] & [mas] & log-flat & power-law & ["] \\
\midrule
384307763771387776 & 2789-1319-1 & 00:00:10.11 & +41:51:42.22 & -7.34 & -4.57 & 1.21 & 384313742365860992 & 2789-1122-1 & 00:00:46.53 & +42:00:37.43 & -6.46 & -4.31 & 1.72 & 1.0 & 0.0 & 672.0 \\
2765217352391138688 & 594-134-1 & 00:00:13.62 & +09:47:56.43 & 26.82 & -2.38 & 2.8 & 2765236731283414784 & 594-81-1 & 00:00:08.28 & +09:56:05.50 & 27.47 & -1.72 & 2.8 & 0.0 & 0.9973 & 495.4 \\
2305900664854805632 & 7526-320-1 & 00:00:20.55 & -41:02:27.77 & 48.7 & -8.96 & 6.28 & 2305905578297389312 & 7526-515-1 & 00:01:39.22 & -40:53:26.62 & 49.46 & -7.99 & 5.31 & 0.9998 & 1.0 & 1042.4 \\
429342935412328704 & 4014-3283-1 & 00:00:26.06 & +60:25:39.51 & 28.74 & 11.84 & 10.21 & 429343004131805568 & 4014-1005-1 & 00:00:24.83 & +60:25:31.28 & 28.11 & 11.49 & 10.28 & 1.0 & 1.0 & 12.3 \\
2880169678167998336 & 2271-911-1 & 00:00:49.66 & +36:46:48.02 & -24.59 & -22.27 & 6.82 & 2880169884326428672 & 2271-1988-1 & 00:00:48.62 & +36:46:38.94 & -25.78 & -23.14 & 7.09 & 1.0 & 1.0 & 15.4 \\
393882207946562176 & 3254-569-1 & 00:00:59.43 & +49:50:54.46 & 6.05 & 4.77 & 2.75 & 393882311025775616 & 3254-1352-1 & 00:00:58.97 & +49:52:02.10 & 7.02 & 4.13 & 2.09 & 0.9979 & 0.9864 & 68.7 \\
4702974002115657472 & 9137-1774-1 & 00:01:17.61 & -70:11:12.28 & 65.45 & -12.93 & 4.67 & 4702974002115657600 & 9137-1708-1 & 00:01:22.50 & -70:11:26.93 & 65.61 & -13.54 & 4.61 & 1.0 & 1.0 & 28.2 \\
531236430303749120 & 4298-598-1 & 00:01:18.66 & +70:55:44.07 & -5.31 & -7.7 & 3.19 & 531236670821917568 & 4298-574-1 & 00:01:03.45 & +70:55:47.55 & -4.8 & -6.92 & 3.44 & 0.9988 & 0.9927 & 74.6 \\
2738660435728687872 & 1-1016-1 & 00:01:42.40 & +01:47:12.51 & -19.21 & -26.34 & 2.66 & 2738660882405286272 & 1-168-1 & 00:01:41.56 & +01:48:24.41 & -19.67 & -25.14 & 2.26 & 0.9998 & 0.8704 & 73.0 \\
2848236321444116352 & 1729-1129-1 & 00:02:01.76 & +23:46:51.13 & 26.99 & 5.65 & 5.04 & 2848423994335483264 & 1729-1118-1 & 00:01:59.86 & +23:47:01.77 & 24.44 & 6.34 & 5.17 & 1.0 & 1.0 & 28.1 \\
\bottomrule
\end{tabular}
}
\end{sidewaystable}

\section{Results}
\label{sec:results}

\subsection{The TGAS Catalog of Wide Binaries}
\label{sec:catalog_describe}
We define wide binaries as systems with a posterior probability above 99\% for either the log-flat or the power-law prior model, and our final catalog of 6196 pairs is the union of these two samples. The posterior probabilities may differ substantially between the  models, particularly at large $s$ where the differences in the prior distributions is strongest.

We note that the posterior probabilities are derived from a calibrated and validated but ultimately imperfect model; pairs with $P(C_2 \given \vec{x}_i, \vec{x}_j) >$ 99\% may still be contaminating objects. Indeed, at separations $s$ $\gtrsim 4\times 10^4$ AU, we suspect most pairs are random alignments regardless of their probability. We discuss this possibility further in Section \ref{sec:1pc_binaries}. Nevertheless, we include these systems in the catalog:  they define the locus of random alignments, and there are likely to be genuine pairs within this sample that can be extracted with a more focused analysis.

We present our catalog in Table \ref{tab:catalog}. When individual stars match to multiple companions, each pair is listed as a separate binary; we do not use these systems when determining the wide binary orbital separation distribution (Sections \ref{sec:orb_sep} and \ref{sec:s_model}). We provide the {\it Gaia} source IDs, Tycho-2 source IDs, positions, proper motions, and parallaxes of the components of each system, the posterior probabilities of each binary being composed of a genuine pair, and their angular separation.

\subsection{The Orbital Separation Distribution of Wide Binaries}
\label{sec:orb_sep}

\begin{figure}
\begin{center}
\includegraphics[width=1.04\columnwidth]{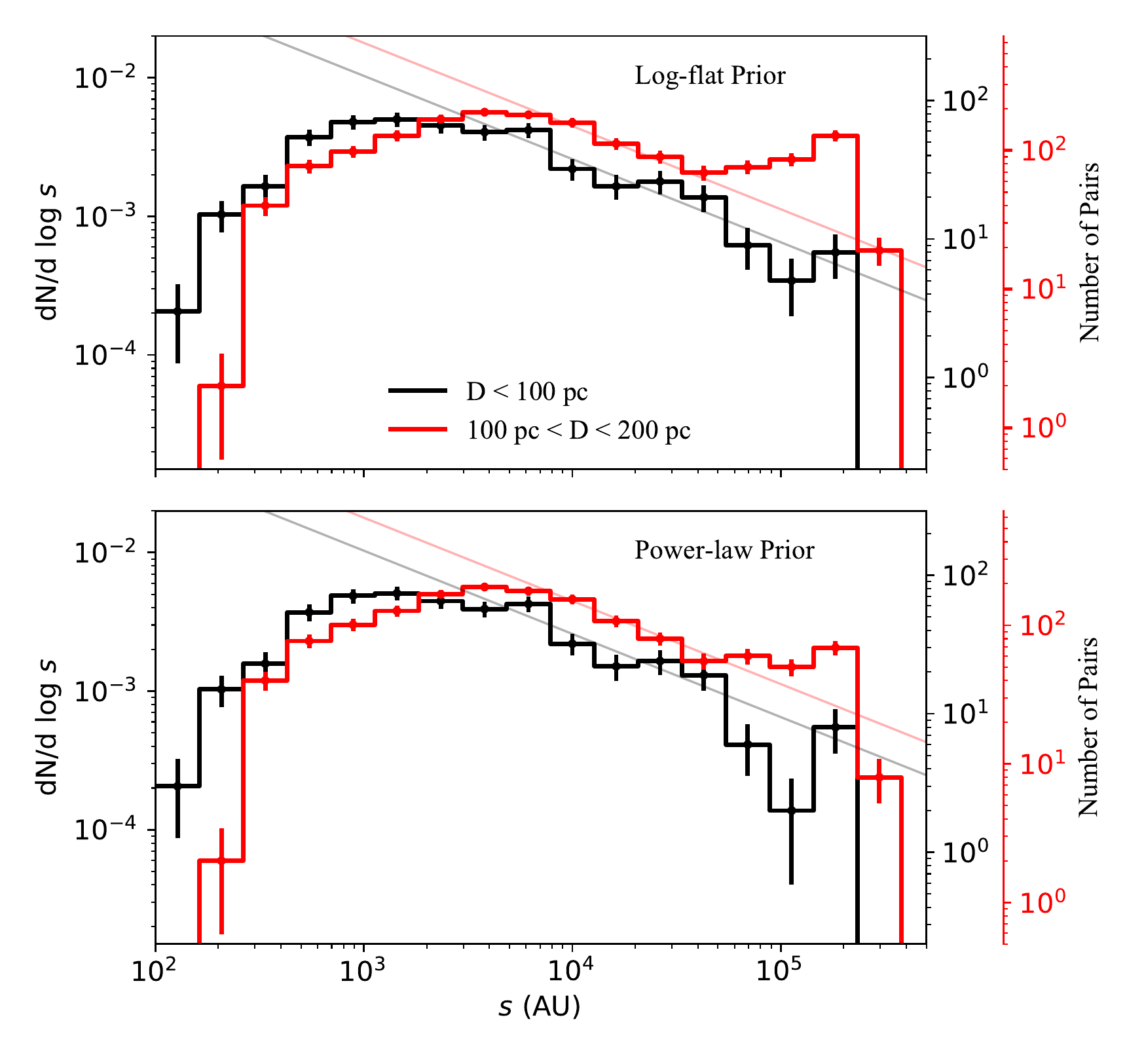}
\caption{ The $s$ distribution of our binary samples identified using both the log-flat prior (top panel) and the power-law prior (bottom panel). The left axis shows these distributions normalized to the total number of TGAS stars within the distance ranges listed, while the right axis shows the overall numbers. Background lines show the power-law $P(s)\propto s^{-1.6}$, which decreases at larger $s$. These lines are not fits to the distribution, but rather the normalization is scaled arbitrarily to compare the distribution to the power-law model at large separations. Regardless of the assumed prior, there is a drop off in the number of binaries with $s\ \gapprox\ 3\times10^3$ AU. The excess at 10$^5$ AU is due to random alignments, as can be seen in the right panels of Figure \ref{fig:TGAS_dist_s_theta}; contamination is expected as the distance increases, which is why the excess observed in the sample with 10$^2 < D < 2\times10^2$ pc (red) is not observed in pairs within 10$^2$ pc (black). }
\label{fig:TGAS_s_distribution}
\end{center}
\end{figure}

In Figure \ref{fig:TGAS_s_distribution}, we show the distribution of projected physical separations for subsets of pairs within two different distance limits for our log-flat and power-law prior models. These distributions are normalized to the total number of TGAS stars within the listed distance ranges. The small angle limit\footnote{Although our catalog of wide binaries contains pairs with separations as small as 2\asec, Figure \ref{fig:TGAS_theta_distribution} shows there is a substantial observational bias in effect at $\theta < 10$\asec.} for our sample leads to an excess of systems at $s\leq2\times10^2$ AU within $10^2$ pc (black) compared with our more distant sample (red). Importantly, the more distant binaries show an excess at $s \approx 10^5$~AU that is unseen in the closer sample at distances $\leq10^2$ pc. This excess at large separations has been observed in other wide binary samples \citep{dhital10,oelkers17} and suggested to be a characteristic of the Galactic wide binary population.

Were this increase a genuine characteristic of wide binaries, it should be seen in the distributions shown in Figure \ref{fig:TGAS_s_distribution} of both the samples within 10$^2$ pc and that at distances between 10$^2$$-$$2\times10^2$ pc. Comparison between Figures \ref{fig:TGAS_dist_s_theta_false} and \ref{fig:TGAS_dist_s_theta} as well as the RV matching shown in Figure \ref{fig:contamination} demonstrates that this increase occurs at the separation where random alignments dominate the sample. We conclude that the relative flattening and subsequent apparent excess of the distribution at $s\approx 4\times 10^4$ AU in systems with $10^2$~<~D~<~$ 2\times10^2$~pc corresponds to the transition from genuine binaries to random alignments seen in the right panels of Figure~\ref{fig:TGAS_dist_s_theta}. In particular, our data imply that the previously reported bimodality within separations of 1 pc in samples of wide binaries \citep{dhital10,oelkers17} is most likely due to contamination from random alignments.

Additionally, the distribution of projected separations of our wide binaries, regardless of our assumed orbital separation prior, is in conflict with that found by \citet{oh2016}, who report an {\it increasing} distribution of co-moving pairs with $s$ ranging from a few 10$^3$~AU to several pc. However, our sample is in agreement with CG04 and \citet{lepine07}, both of who find samples of wide binaries decreasing at separations larger than a few 10$^3$ AU. A major advantage of our sample over that of \citet{oh2016} is that we explicitly identify and estimate the contamination fraction; after removing potential contamination our cleaned sample is nearly two orders of magnitude larger than that of \citet{oh2016} for pairs with projected separations smaller than $4\times10^4$ AU. If these authors introduced a strong selection function that distorts the observed distribution of stellar pairs, the difference between their sample and ours can be resolved.

\citet{oh2016} further identify a set of co-moving pairs with $s\gtrsim1$ pc, speculating that these may be the unbound remnants of ionized  binaries whose components disperse in phase space during their orbital evolution. As \citet{jiang10} demonstrate, such co-moving pairs may be prevalent in the nearby Galaxy. We discuss the nature of the population of pairs with separations $s\gtrsim1$  pc in Section \ref{sec:1pc_binaries}; for now we note that the distribution of wide binaries reported by \citet{oh2016} increases in the same way as a population of purely random pairs, a  sign of potential contamination.

Our sample of wide binaries shows a sharp cutoff at separations of a few 10$^5$ AU. This corresponds to the limits imposed by our binary model defined in Section \ref{sec:challenges}: based on the expectation that (gravitationally bound) binaries do not survive for long at orbital separations beyond the Galactic tidal limit, our model has a strict upper limit that $a\leq$ 1 pc. As noted earlier, binaries with separations $>$1 pc may exist and indeed have been identified \citep[e.g., CG04;][]{shaya11}. However, our results indicate that correctly identifying such widely separated systems as genuinely associated pairs is very challenging.

The dotted lines in Figure \ref{fig:TGAS_s_distribution} correspond to the power-law distribution observed by CG04 and \citet{lepine07} with $P(s) \propto s^{-1.6}$. These lines are scaled vertically to compare the distribution to the power-law model at large separations. Random alignments artificially increase the distribution at large $s$ for the pairs with $10^2 < D < 2\times10^2$~pc. Because the $D<10^2$ pc sample has less contamination at larger $s$, the power-law provides a reasonable model from $\approx$$5\times 10^3$ to 10$^5$ AU. In the following section, we demonstrate that this decrease at large $s$ cannot be due to observational biases in detecting binaries with large $\theta$. The distribution of $s$ is clearly inconsistent with {\"O}pik's Law for $s\ \gapprox\ 10^4$ AU, and it is not bimodal for $s\lesssim$ 1 pc.

\subsection{Comparison to Models}
\label{sec:s_model}
To go beyond our by-eye comparison to the power-law separation distribution shown in Figure \ref{fig:TGAS_s_distribution}, we wish to use our full catalog of 6196 systems with posterior probabilities above 99\%. The interdependence of distance, projected separation, and angular separation of these binaries shown in the right panels of Figure \ref{fig:TGAS_dist_s_theta} demonstrates the difficulty in deriving the underlying distribution of physical separations. Contamination at large $s$, bias introduced by the distance distribution of stars in TGAS, and the angular resolution limit of TGAS at small $s$ may all cause inaccuracies when determining the distribution of $s$. Our approach is to generate a sample of synthetic wide binaries from an arbitrary binary population model, convert those binaries to an angular separation distribution, $P(\theta)$, accounting for the angular separation limits and distance dependence of TGAS. In this way, we are calculating $P(\theta)$ as predicted by the models to directly compare to subsets of our sample of binaries.

\begin{figure*}
\begin{center}
\includegraphics[width=0.85\textwidth]{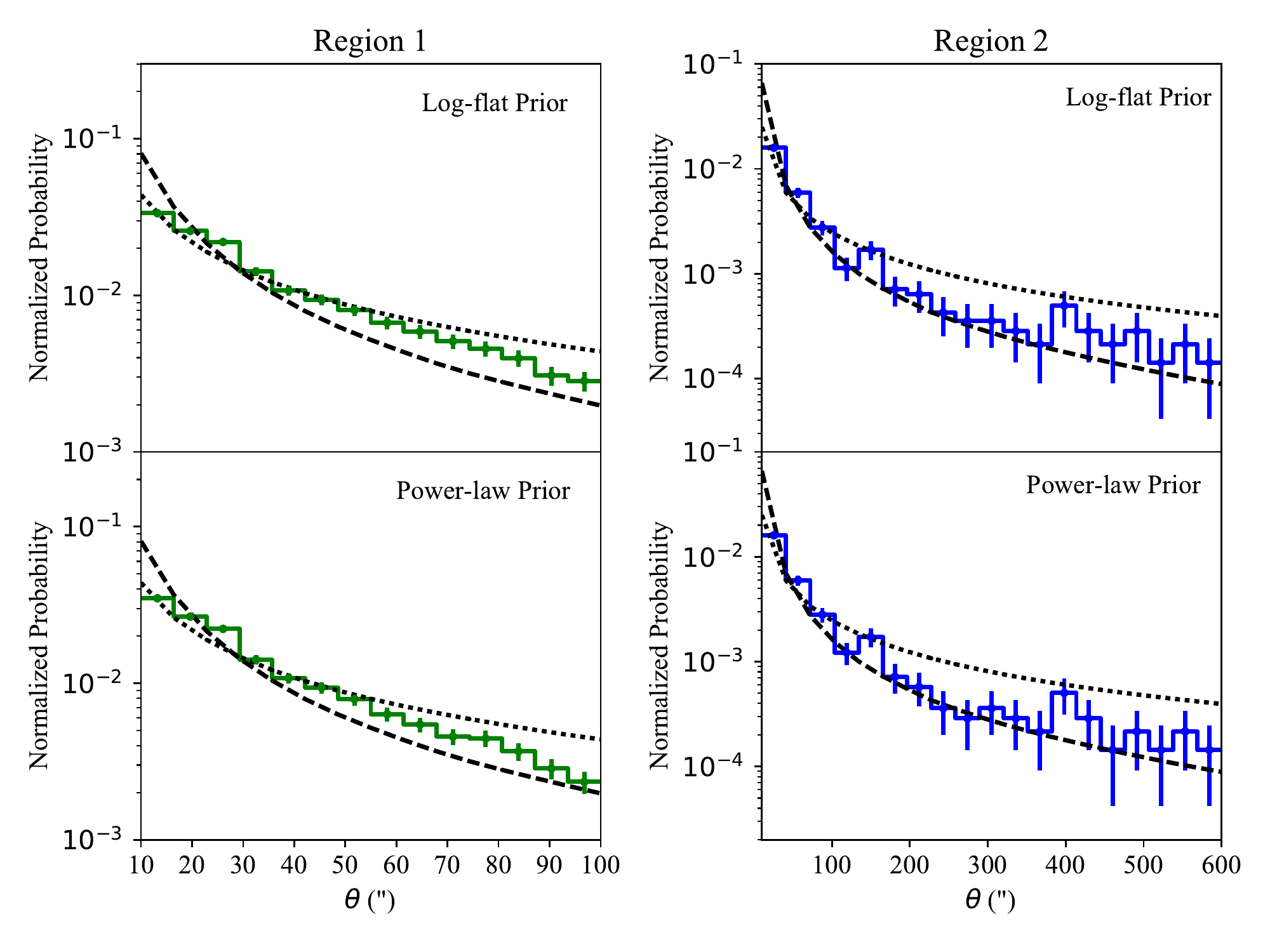}
\caption{$P(\theta)$ for two models: one based on \"{O}pik's Law, in which $a$ is selected randomly from a log-flat distribution (dotted lines), and one with a power-law distribution (dashed lines). $P(\theta)$ in both cases is calculated using the method outlined in Section~\ref{sec:s_model}. We compare these models to the two different samples of pairs we generate, one using the log-flat prior (top panels) and one using the power-law prior (bottom panels). The left and right columns of panels are computed for Regions 1 and 2, respectively, shown in Figure~\ref{fig:TGAS_dist_s_theta} and described in Section~\ref{sec:orb_sep}. In every panel, we compare our samples to the two model distributions. Regardless of the prior distribution assumed, our model robustly indicates that at larger separations, $\theta$ corresponds to a power-law steeper than flat and consistent with $P(s)\propto s^{-1.6}$, while at smaller separations the distribution flattens out and is consistent with a log-flat distribution. }
\label{fig:TGAS_theta_dist}
\end{center}
\end{figure*}

We restrict ourselves to Regions 1 and 2, defined in Figure \ref{fig:TGAS_dist_s_theta}, to calculate $P(\theta)$ expected from the model by integrating over the two oblique regions in $D$-$s$ space and then factoring terms based on independence:
\begin{eqnarray}
P(\theta) &=& \mathcal{Z} \int_0^{D_{\rm max}} \int_{s_{\rm min}}^{s_{\rm max}} \dd D\ \dd s\ P(\theta, D, s) \nonumber \\ 
&=& \mathcal{Z} \int_0^{D_{\rm max}} \int_{s_{\rm min}}^{s_{\rm max}} \dd D\ \dd s\ P(\theta \given D, s)\ P(D)\ P(s),
\label{eq:P_theta_marginalize}
\end{eqnarray}
where $\mathcal{Z}$ is an integration constant to normalize the probability and $s_{\rm min}$ and $s_{\rm max}$ are the minimum and maximum projected separations, respectively, to which we are sensitive. For a given distance, these are limited by our angular separation constraints: at the close end by the angular sensitivity of the Tycho-2 catalog and at the distant end by our search radius of 100\asec\ for Region 1 and 10\amin\ for Region 2.

Since $\theta$ can be expressed as a function of $D$ and $s$, Equation \ref{eq:P_theta_marginalize} can be reduced by substituting a Dirac delta function for $P(\theta \given D, s)$:
\begin{equation}
P(\theta) = \mathcal{Z} \int_0^{D_{\rm max}} \int_{s_{\rm min}}^{s_{\rm max}} \dd D\ \dd s\ \delta\left[ \theta - \frac{s}{D} \right]\ P(D)\ P(s). 
\label{eq:P_theta_delta}
\end{equation}
The delta function reduces the inner integral over $s$. After dividing by the Jacobian term, $1/D$, which comes from the derivative of the argument of the delta function, Equation \ref{eq:P_theta_delta} becomes:
\begin{equation}
P(\theta) = \mathcal{Z} \int_0^{D_{\rm max}} \dd D\ D\ P(D)\ P(s^*),
\label{eq:P_theta_final}
\end{equation}
where $s^* = \theta D$. $P(D)$ is the distance distribution of our sample of binaries, and $P(s^*)$ depends on the specific binary model used. We determine $P(D)$ empirically by generating a KDE from the subsample of our  binaries with $5\times 10^3 < s < 5\times 10^4$ AU. This sample should be free of biases imposed by the TGAS angular separation limits  and of contamination from random alignments at large $s$.

To generate $P(s)$, we create 10$^5$ binaries with randomly chosen orbital parameters and calculate the projected physical separation in the same way as described in Section~\ref{sec:binary_likelihood}. A thermal eccentricity distribution is assumed. Again, we generate $P(s)$ for two models, one with a log-flat distribution for the orbital separation in the range 10 \Rsun$-$1 pc, and one with a power-law of index  $-$1.6. We approximate each of these distributions of 10$^5$ random binaries with gaussian KDEs to generate $P(s)$.  

The integral in Equation \ref{eq:P_theta_final} is normalized over the entire range of $\theta$. To obtain $P(\theta)$ over the range to which we are sensitive, we need to calculate the integration constant $\mathcal{Z}$:
\begin{equation}
\mathcal{Z}^{-1} = \int_{\theta_{\rm min}}^{\theta_{\rm max}} \dd \theta\ P(\theta),
\label{eq:Z}
\end{equation}
where $\theta_{\rm min}$ and $\theta_{\rm max}$ are the angular separation limits defining the region to which we are sensitive. To avoid incompleteness at small separations and random alignments at large separations, we set $\theta_{\rm min}$ to 10\asec\ and $\theta_{\rm max}$ to 10\amin\ for Region 1 and $\theta_{\rm max}$ to 100\asec\ for Region 2.

We compare the actual $P(\theta)$ for binaries in our two samples to these distributions. For both the log-flat and power-law distributions, we find $P(\theta)$ by numerically calculating the integral in Equation \ref{eq:P_theta_final}. Each panel in Figure \ref{fig:TGAS_theta_dist} compares $P(\theta)$ for each of these two models. The top row of panels shows the distributions for our sample using the power-law model as our binary population prior, while the bottom panel shows that for the log-flat  prior. The two columns compare the distributions for Region 1 extending to 10\amin\ and 100 pc and Region 2 extending to 100\asec\ and 500 pc.

The left panels in Figure \ref{fig:TGAS_theta_dist}, calculated using those binaries in Region 1, focus on the distributions of pairs with small $\theta$. Although the two samples are very similar at these separations, the number of binaries in our sample allow us to discern between the two models. For $\theta \ \lapprox\ 80$\asec, the population starts to have a shallower slope, closer to the log-flat distribution. This indicates that the turnover observed at separations of $\approx$$5\times 10^3$ AU in Figure~\ref{fig:TGAS_s_distribution} is genuine.

The right two panels, calculated for binaries in Region 2, show that, regardless of the binary population model assumed, at separations out to 300\asec\ the observed population matches the power-law distribution. Beyond 300\asec, the distribution deviates slightly, which may be due to the increased importance of random alignments (however, the uncertainties at separations beyond 300\asec\ are large). At smaller angular separations, the two models overlap.

Importantly, our results are identical regardless of the priors we use: both models show that $P(s)$ should be roughly log-flat at small separations and scale with roughly $s^{-1.6}$ at large separations. These results are consistent with the results of CG04 and \citet{lepine07}, both of whom, using independent samples, find a flattening of the distribution at small $s$. For instance, \citet{lepine07} determines that $P(s) \propto s^{-1}$ for $s<2500$ AU while at larger separations, $P(s) \propto s^{-1.6}$; for $\theta < 10\asec$, the log-flat and power-law distributions diverge further. Future {\it Gaia} data releases will be sensitive to these differences.

\section{Discussion}
\label{sec:discussion}

\subsection{Comparison with Pre-{\it Gaia} Wide Binary Samples}
\label{sec:compare_catalogs}
Searches of proper-motion catalogs for wide binaries have used different astrometric catalogs and were therefore sensitive to different samples of wide binaries. We briefly summarize the work of several of these groups before comparing their samples to that identified in this work.

CG04 searched the rNLTT catalog, a reanalysis of stars in the NLTT catalog \citep{salim03}, to identify wide binaries among common proper motion pairs. The stars in this catalog extend to $V\approx19$ mag; the relative proper motion sensitivity is $\approx$3 mas yr$^{-1}$.

\citet{lepine07} searched for common proper motion companions to {\it Hipparcos} stars in the LPSM-North catalog, a catalog identified using a reanalysis of the stars in the Digitized Sky Survey and containing stars with $\mu>0.15\asec$~yr$^{-1}$ \citep{lepine05}. Since the primary stars are found in {\it Hipparcos} data, this catalog contains the brightest and nearest stars.  

\begin{figure*}
\begin{center}
\includegraphics[width=1.01\textwidth]{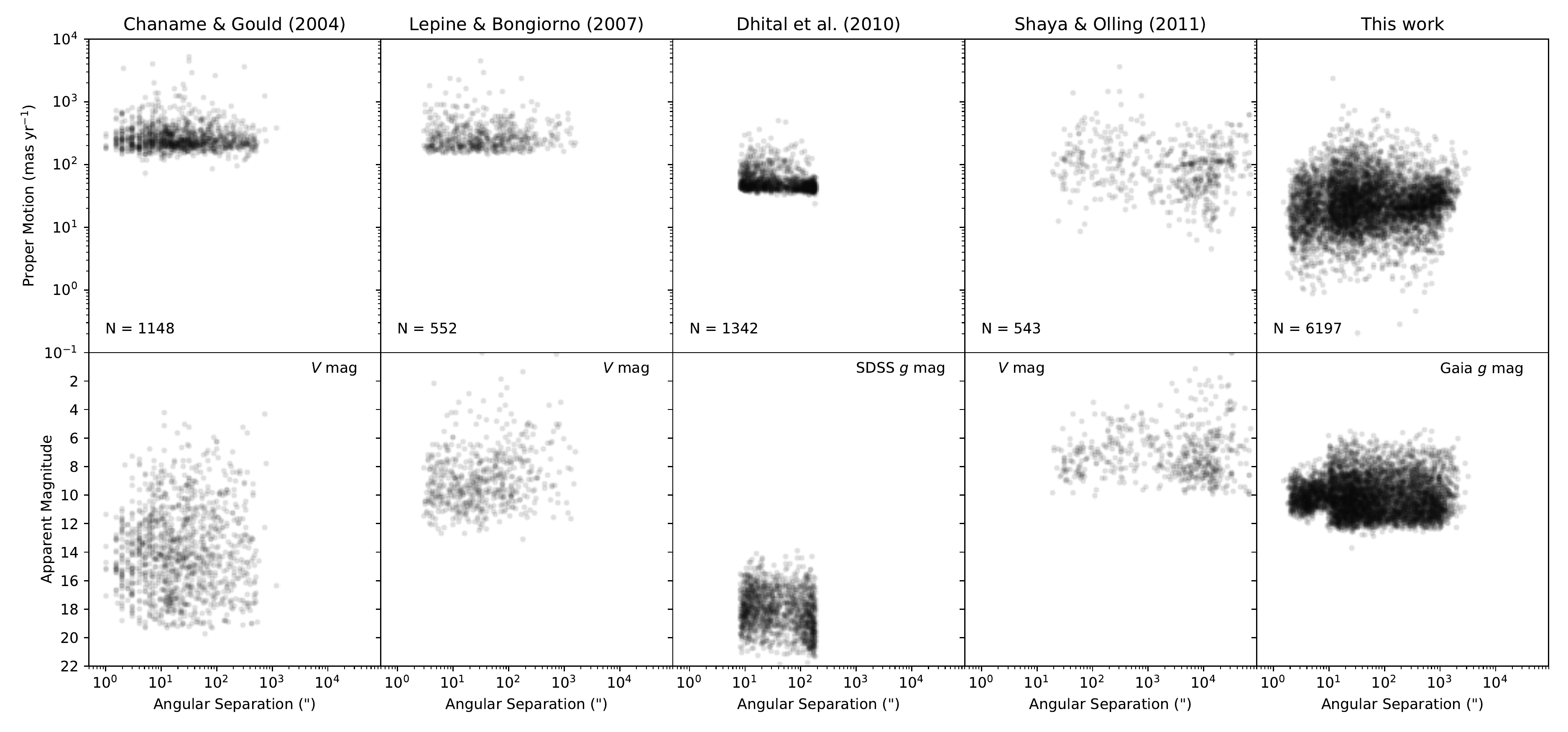}
\caption{ The two panels in each column show the distribution of angular separations, proper motions, and magnitudes for four samples of wide binaries in the literature and for the sample identified in this work. While previous catalogs were limited to nearby pairs with large proper motions, the TGAS astrometric precision allows us to identify binaries with proper motions down to a few mas~yr$^{-1}$. However, the photometric limitations of the Tycho-2 catalog result in the pairs identified in this work being relatively brighter than previously identified pairs \citep[with the exception of the catalog of][which is based on stars with {\it Hipparcos} astrometry]{shaya11}. Future {\it Gaia} data releases should allow us to extend this sample to fainter magnitudes. Note that the magnitudes shown are in the bands provided in the respective catalogs.}
\label{fig:compare_catalog}
\end{center}
\end{figure*}

\citet{dhital10} searched SDSS for common proper motion pairs containing low-mass stars. Proper motions are provided by the SDSS team by comparing to photometric plates from the USNO-B and POSS-II surveys \citep{munn04}. The digital nature of SDSS and the long time baseline results in proper motions with precisions $\approx$few mas yr$^{-1}$ for stars as faint as $g\approx20$ mag. However, this sample is limited by several factors: source confusion prevents pairs with angular separations $\lesssim$7\asec\ from being identified and SDSS has a bright limit of $g\approx14$ and limited sky coverage.  

Finally, \citet{shaya11} searched the {\it Hipparcos} catalog for wide binaries. This sample includes only the brightest pairs, since the {\it Hipparcos} catalog is limited to $V \lesssim 9$ mag; however, the resulting catalog extends to proper motions as small as a few mas~yr$^{-1}$. \citet{shaya11} were able to identify such slowly moving pairs because their algorithm also matched stars based on {\it Hipparcos} parallaxes.

The left four columns in Figure \ref{fig:compare_catalog} compares the angular separations, magnitudes, and proper motions of pairs identified in the catalogs of CG04, \citet{lepine07}, \citet{dhital10}, and \citet{shaya11}.\footnote{We only include those pairs identified by \citet{shaya11} with a probability above 50\%.} The right column shows the distributions for the wide binaries identified in this work. The bottom panel in the rightmost column demonstrates that, like for \citet{shaya11} and \citet{lepine07}, our pairs tend to be at brighter magnitudes ({\it Gaia} $g \lesssim 11$ mag), reflecting the relatively bright faintness limit of the Tycho-2 catalog. However, even more so than \citet{shaya11}, we are identifying pairs with smaller proper motions.

The number of pairs in each catalog is provided in the top panel of each column. Our catalog represents an increase in size of a factor of $\approx$5-10 over each of the previous catalogs. At the same time, the CG04 and \citet{dhital10} catalogs shown in Figure~\ref{fig:compare_catalog} indicate that many more pairs exist with fainter magnitudes and large proper motions. Therefore, we can infer that the sample of wide binaries identified in this work represents a small subset of a much larger population that will be identified in future {\it Gaia} data releases.

\subsection{The Impact of Our Assumptions}
\label{sec:assumptions}
We discuss two of the three most important assumptions we made in constructing our catalog: the prior on the parallax and the binary fraction. The other assumption, about the input binary population model, was discussed in Section~\ref{sec:binary_likelihood}. 

\subsubsection{The Prior on the Parallax}
Through their analysis of individual stars, \citet{astraatmadja16} test several prior distributions on parallax. These authors identify a model with an exponentially decreasing density with a scale length of 1.35 kpc as a reasonably accurate description of the parallax distribution. This is the prior we use, and we provide its functional form in Section \ref{sec:binary_likelihood}. 

As discussed by \citet{lutz73} and confirmed for {\it Gaia} data by \citet{astraatmadja16}, the astrometric parallax is a poor measure of distance when its fractional uncertainty, $\sigma_{\varpi}/\varpi$, is $\gapprox$20\%. Stars with small $\sigma_{\varpi}/\varpi$ (which are usually closer by) are less sensitive to the assumed parallax prior. Since the typical TGAS $\sigma_{\varpi} \leq 0.3$~mas, the parallaxes should be accurate for stars within $\lapprox$700 pc, which describes the vast majority of our wide binaries. 

We test this assumption by running our algorithm with an empirically motivated parallax model based on the distribution of observed parallaxes. Specifically, we generate a KDE of a subset of parallaxes in the TGAS catalog to approximate this distribution. Uncertainties will tend to make a population of stars with small parallaxes appear somewhat closer, and therefore to have larger parallaxes, than at their true distances \citep[Lutz-Kelker bias;][]{lutz73}. Our empirical parallax prior is therefore an extreme case. 

After comparing the results of our model to the two different parallax distributions, we find that the samples at separations $\lapprox$few 10$^4$ AU, from which we draw our conclusions about the properties of wide binaries, remain essentially unchanged. Since these very different parallax distributions produce nearly identical wide binary catalogs, we conclude that our results will be unaffected by even large differences in parallax prior probabilities.  However, we note that this may not be true for future {\it Gaia} catalogs with different proper motion, parallax, and magnitude properties.

\subsubsection{The Binary Fraction}
\label{sec:binary_fraction}

Estimates of the binary fraction are a strong function of mass, and range widely, from 44\% for solar-like stars \citep{raghavan10} to $>$80\% for O stars \citep{chini12}. However, these fractions are typically calculated for spectroscopic binaries. For wide binaries, \citet{lepine07} calculated that at least 10\% of nearby {\it Hipparcos} stars have companions at separations $>$10$^3$~AU, and the wide binary fraction may be somewhat larger. Here, we assume that the {\it total} binary fraction $f_{\rm bin}$ is 50\%, which enters into Equation \ref{eq:P_binary_2} as the prior probability on $C_2$. The prior is normalized such that half of these stars, or 25\% overall, have companions with separations $>$10$^{2}$ AU.

In addition, we define $f_{\rm bin}$ to be the fraction of binaries visible to {\it Gaia}, rather than the fraction of all stars in  binaries. Due to the relatively bright faint limit of Tycho-2 photometry, many stars with faint companions cannot be identified in our data. If the wide binary fraction in the Galaxy is 50\%, we would need a somewhat lower value for $f_{\rm bin}$ to account for our inability to detect pairs with faint companions. This bias will be substantially reduced in future {\it Gaia} data releases.

Fortunately, there is some freedom in our model to account for an inaccurate value for $f_{\rm bin}$, as it scales monotonically with our posterior probability $P(C_2 \given \vec{x}_i, \vec{x}_j)$ in Equation \ref{eq:P_binary_2}. If our adopted value for $f_{\rm bin}$ is too high, it will be reflected in the distribution of $P(C_2 \given \vec{x}_i, \vec{x}_j)$ for both genuine pairs and random alignments. In separating genuine binaries from random alignments using our independent RV and random alignment tests, we calibrate the critical value for $P(C_2 \given \vec{x}_i, \vec{x}_j)$ and  account somewhat for our ignorance of $f_{\rm bin}$. Indeed, one of the reasons we find a critical value for $P(C_2 \given \vec{x}_i, \vec{x}_j)$ of 99\% rather than a more intuitive 50\% may be because our assumed value of $f_{\rm bin}$ is too high.

\subsection{Stellar Pairs at pc-scale Separations}
\label{sec:1pc_binaries}

Stellar pairs with $s\gtrsim1$ pc are quickly dissociated by the Galactic tide and are therefore not formally bound. However, these formerly bound pairs, which we consider co-moving pairs rather than  binaries, may remain in similar locations in phase space for many orbits around the Galaxy \citep{jiang10}. \citet{jiang10} find two peaks in the distribution of stellar separations, one well within 1 pc and one at several 100 pc. Importantly, regardless of their input model, these authors find that the trough in the distribution of stellar separations between these two peaks is at $\approx$5 Jacobi radii ($\approx$10 pc in the local Galactic neighborhood). There is a strong motivation to identify stellar pairs at separations $>$1 pc since detection of this trough followed by the second peak in the projected separation distribution of co-moving pairs would provide confirmation of the dissociating effect of the Galactic tide and potentially a powerful probe of Galactic structure.

Previous authors have claimed to identify large numbers of binaries or co-moving stellar pairs at wide separations \citep{shaya11,oh2016,oelkers17}, and our search method identifies candidate binaries at projected separations $\sim$pc. Here we list several reasons that lead us to suspect that a large fraction of these stellar pairs, and by extension potentially those in other samples, are randomly aligned stars.

First, Figure \ref{fig:TGAS_dist_s_theta} shows that these pairs have positions in log $D$$-$log $s$ space coincident with that of our sample of randomly aligned stars, derived from applying our method to a shifted version of the TGAS catalog. Second, the population of these pairs is strongly sensitive to the assumptions of the model. Our log-flat and power-law priors produce very different numbers of candidate pairs with $s\sim 10^5$~AU. This implies that the identification of any individual co-moving pair at these separations is tenuous and model-dependent. Third, Figure \ref{fig:TGAS_s_distribution} shows that, regardless of the underlying model assumptions, the distribution of pairs at $s\sim 10^5$ AU is distance-dependent. This population is seen in the distribution of our sample at distances between 10$^2$ and $2\times10^2$ pc, but not in the distribution within the nearest $10^2$ pc. However, there is no clear reason why genuine co-moving pairs with pc-scale separations might exist at distances beyond $10^2$ pc, but not within $10^2$ pc. Furthermore, in Figure \ref{fig:contamination} we show the difference in RVs of pairs with RAVE data as a function of $s$. Our results indicate that stellar pairs with $s \gtrsim 4\times 10^4$~AU identified from TGAS, even with high posterior probabilities, typically have inconsistent RVs.  

Nevertheless, co-moving pairs at these wide separations may exist, and if they do, our method and others may classify them as candidates. For instance, \citet{oh2016} identify a sample of co-moving pairs in TGAS with $s\gtrsim 1$ pc and demonstrate using matching with RAVE that the majority with a separation $<$few pc have consistent RVs. However, at separations $>$few pc, the RVs quickly become more consistent with random alignments. Furthermore, these authors find a distribution of projected separations linearly increasing out to $s\approx10$ pc, the separation at which \citet{jiang10} predict a trough in the distribution of genuine co-moving  pairs. This increasing distribution with projected separation can be readily explained if the \citet{oh2016} sample is dominated at these separations by randomly aligned stellar pairs: the number of random alignments increases with phase space volume, and therefore linearly with projected separation.

\citet{oelkers17} find several thousand co-moving pairs at separations $>$1 pc. As these authors estimate contamination based on an assumed Galactic model rather than a data-driven, independent method such as matching RVs, we suggest that at separations $\gapprox 4\times 10^4$ AU random alignments may be a significant source of contamination in their catalog, similar to what we see in our data.

Identifying wide stellar pairs that cluster in phase space, as is done by \cite{oh2016}, may be effective for detecting co-moving stellar conglomerations such as triple or higher multiplicity systems, open clusters, OB associations, and moving groups. However, the analysis we perform indicates that at separations beyond $\sim$1 pc it becomes increasingly difficult to separate isolated co-moving pairs from randomly aligned stars. We suggest that future searches for such pairs include careful accounting for contamination and, if possible, validation with RVs.

\subsection{Using this Catalog}
\label{sec:using_this_catalog}

Before using the catalog provided in Table \ref{tab:catalog} there are a number of caveats to consider. First, the immense majority of pairs with posterior probabilities $>$90\% are likely to be random alignments, and our results are derived from only those systems with posterior probabilities $>$99\%. Nevertheless, there may be some contaminating random alignments in this sample. As discussed in Sections \ref{sec:dist_s_theta} and \ref{sec:RV}, randomly aligned pairs typically have $s > 4\times 10^4$~AU, while pairs with $s < 4\times 10^4$~AU are mostly genuine. 

However, the transition between genuine pairs and random alignments is not sharp; genuine pairs with $s> 4\times10^4$~AU likely exist in our catalog, as may random alignments with very small separations. In particular, Figure~\ref{fig:TGAS_dist_s_theta_false} shows that random alignments may have angular separations as small as 1\asec. To obtain a large and pure sample, we suggest using the subset of our candidate pairs in which the power-law posterior probability is above 99\% and $s<4\times 10^4$~AU. Based on the tests described in Section \ref{sec:RV}, we expect the contamination of such a sample to be $\lapprox$5\%.

Contamination due to random alignments becomes even less problematic for the candidate pairs falling within Regions 1 and 2 in Figure~\ref{fig:TGAS_dist_s_theta}. Contamination is nearly non-existent for pairs in Region 2 in particular. However, these data are subject to important observational biases, particularly in distance. When deriving sample statistics where sensitivity to distance may be relevant, this bias should be taken into account. In Section \ref{sec:s_model}, we describe one method for taking this into account using a Bayesian formalism.

Although we do not expect any bias in terms of $\theta$ to exist at separations larger than 10\asec, Figure \ref{fig:TGAS_theta_distribution} shows a marked decrease in the number of stellar pairs with $\theta < 10$\asec.

\section{Conclusions}
\label{sec:conclusions}

\subsection{Summary of Results}

We present a method to identify wide binaries by matching pairs of stars based on their positions, proper motions, and parallaxes. We make no effort to identify co-moving stars in stellar clusters, moving groups, or higher multiplicity systems. We use a fully Bayesian method to discern between randomly aligned double stars and genuine binaries. Uncertainties on the astrometric quantities as well as their covariances are properly included in our analysis. We argue (and confirm using our resulting sample) that {\it Gaia}'s astrometric precision is fine enough to identify small differences in the proper motions of stars in wide binaries due to orbital velocity. We account for these slight velocity differences in our search algorithm. Furthermore, our method makes no assumptions about the structure of the Galaxy, and we calibrate it using independent RV observations. 

As a test, in Appendix \ref{sec:rNLTT} we apply our algorithm to the subset of the revised NLTT catalog with parallaxes measured by {\it Hipparcos} \citep{gould03}. \citet{chaname04}, whose search algorithm did not include parallax, identify 44 wide binaries in this sample. We re-detect these 44 binaries and find an additional 72 new wide binaries. This test demonstrates both the discerning power of including parallax as a matching criterion and the success of our method at parsing catalogs to identify wide binaries.

After applying our method to the 2$\times 10^6$ stars in the TGAS catalog, we identify 6196 wide binaries with a posterior probability above 99\% in at least one of the catalogs produced using our two prior probability distributions for the orbital separation. Matching stars in our pairs with the RAVE catalog provides a subsample of our catalog that have RVs for both components. These RVs provide a test for our method: widely separated but gravitationally bound binaries should have components with matching RVs (modulo the presence of hierarchical triples). This test indicates that our catalog has a high fidelity and is largely free of unassociated, randomly aligned pairs.

When compared to catalogs of previously identified wide binaries, the pairs in this catalog are typically brighter, due to the magnitude limits of the Tycho-2 catalog. Nevertheless, our pairs extend to much smaller proper motions; where previous wide binary catalogs only include binaries with large proper motions, our sample contains binaries with proper motions as small as a few mas yr$^{-1}$.

Our sample suffers from strong biases at both small separations, due to the angular separation limits of the Tycho-2 catalog, and at large angular separations, where contamination from randomly aligned pairs dominates the sample. Nevertheless, we define a method to compare an arbitrary distribution of binary separations with our biased sample.

Our sample indicates that at large physical separations, the distribution of binaries follows roughly a power-law with an index of $-$1.6, in agreement with the wide binary distributions identified by both \citet{chaname04} and \citet{lepine07}. At smaller separations, the distributions flatten to a power-law index closer to $-$1 (\"{O}pik's Law). Importantly, these results are robust, as they are independent of whether our model assumes a log-flat or a power-law distribution as a prior on the orbital separation. A larger sample at a wider range of angular separations is needed to better determine the exact shape of this distribution.

Finally, we do not find a bimodality in the separations, as reported by previous authors \citep{dhital10,oelkers17}. Using RVs and comparing our catalog to a sample of random alignments, we demonstrate that the majority of candidate pairs identified by our method with $s \gtrsim 4\times 10^4$~AU are composed of randomly aligned, unassociated stars. We nevertheless include these pairs in our catalog, since a more focused analysis may be able to extract genuine binaries and co-moving pairs from this sample.

Future {\it Gaia} data releases will include RVs for stars brighter than $\approx$15 mag. Such measurements allow an important additional constraint on the binarity of widely separated pairs, potentially  identifying wide binaries by matching their positions in full six-dimensional phase space. However, stars with matching parallaxes and proper motions but mismatched RVs may still be associated: the pair may actually be a hierarchical triple, with one star in the pair a close, unresolved binary. Engaging in RV follow-up of matched pairs with mismatched RVs may prove to be a unique and efficient method for identifying a population of hierarchical triple systems.

\subsection{Application to Future {\it Gaia} Data Releases}
\label{sec:future_gaia}

Our method is designed to take advantage of {\it Gaia}'s precise astrometry while remaining scalable to larger data sets. Our current algorithm has a computational cost of a few 100 CPU hours for the 2$\times 10^6$ stars in the TGAS catalog. With modest optimization, this algorithm should be applicable to the 10$^9$ stars expected in the upcoming {\it Gaia} data releases. 

Given the magnitude limits of the first {\it Gaia} data release, the pairs identified here are likely to have mass ratios close to unity. We therefore postpone any discussion of population statistics in terms of stellar masses for this sample. Upcoming data releases will extend the sample of stars with five-dimensional astrometry to $V\approx20$. Using the yields from previous searches for wide binaries as an indicator, we estimate that searches through these future {\it Gaia} data sets will produce 10$^4$$-$10$^5$ wide binaries. This population will be large enough to identify robustly the underlying population of wide binaries as a function of mass and ultimately test various models for the formation of wide binaries.

\section*{Acknowledgements}
We thank the anonymous referee for helpful suggestions which greatly improved the manuscript. J.C. thanks Claudia Aguilera-G{\'o}mez for introducing him to TOPCAT, which was extremely useful for working with the large TGAS and RAVE databases. We used the python module {\tt corner} \citep{dfm_corner} to produce several figures in this manuscript. This research made use of Astropy, a community-developed core Python package for Astronomy \citep{astropy}. J.J.A. acknowledges funding from the European Research Council under the European Union's Seventh Framework Programme (FP/2007-2013)/ERC Grant Agreement n. 617001. J.C. acknowledges support from the Chilean Ministry for the Economy, Development, and Tourism's Programa Iniciativa Cient\'{i}fica Milenio, through grant IC120009 awarded to the Millenium Institute of Astrophysics (MAS), from PFB-06 Centro de Astronomia y Tecnologias Afines, and from Proyecto FONDECYT Regular 1130373. This work has made use of data from the European Space Agency (ESA) mission {\it Gaia} (\url{http://www.cosmos.esa.int/gaia}), processed by the {\it Gaia} Data Processing and Analysis Consortium (DPAC, \url{http://www.cosmos.esa.int/web/gaia/dpac/consortium}). Funding for the DPAC has been provided by national institutions, in particular the institutions participating in the {\it Gaia} Multilateral Agreement. We acknowledge use of the Metropolis HPC Facility at the CCQCN Center of the University of Crete, supported by the European Union Seventh Framework Programme (FP7-REGPOT-2012-2013-1) under grant agreement no. 316165.


\bibliographystyle{mnras}
\bibliography{references} 


\appendix
\section{Testing our Method with the Revised NLTT Catalog} \label{sec:rNLTT}

Before applying our method to the TGAS catalog, we tested it against a set of wide binaries already identified in the literature from an astrometric catalog: the rNLTT catalog \citep{gould03, salim03}, which contains 36,085 entries with a typical $\mu$ uncertainty of several mas~yr$^{-1}$. The stars in rNLTT are mostly at latitudes above and below the Galactic Plane and were selected based on their large proper motions. Because of this low density in both position and proper motion space, the probability of randomly aligned pairs is substantially smaller in rNLTT than in TGAS. 

Since our method requires parallaxes, we tested it only on the 8288 rNLTT stars with {\it Hipparcos} parallaxes. We found 213 pairs with a posterior probability $>$99\% of being a genuine binary. However, a fraction of these pairs have both components of the binary matched to the same {\it Hipparcos} star; after removing these, we identified 116 wide binaries in the rNLTT catalog.

\begin{figure}
\begin{center}
\includegraphics[width=1.\columnwidth]{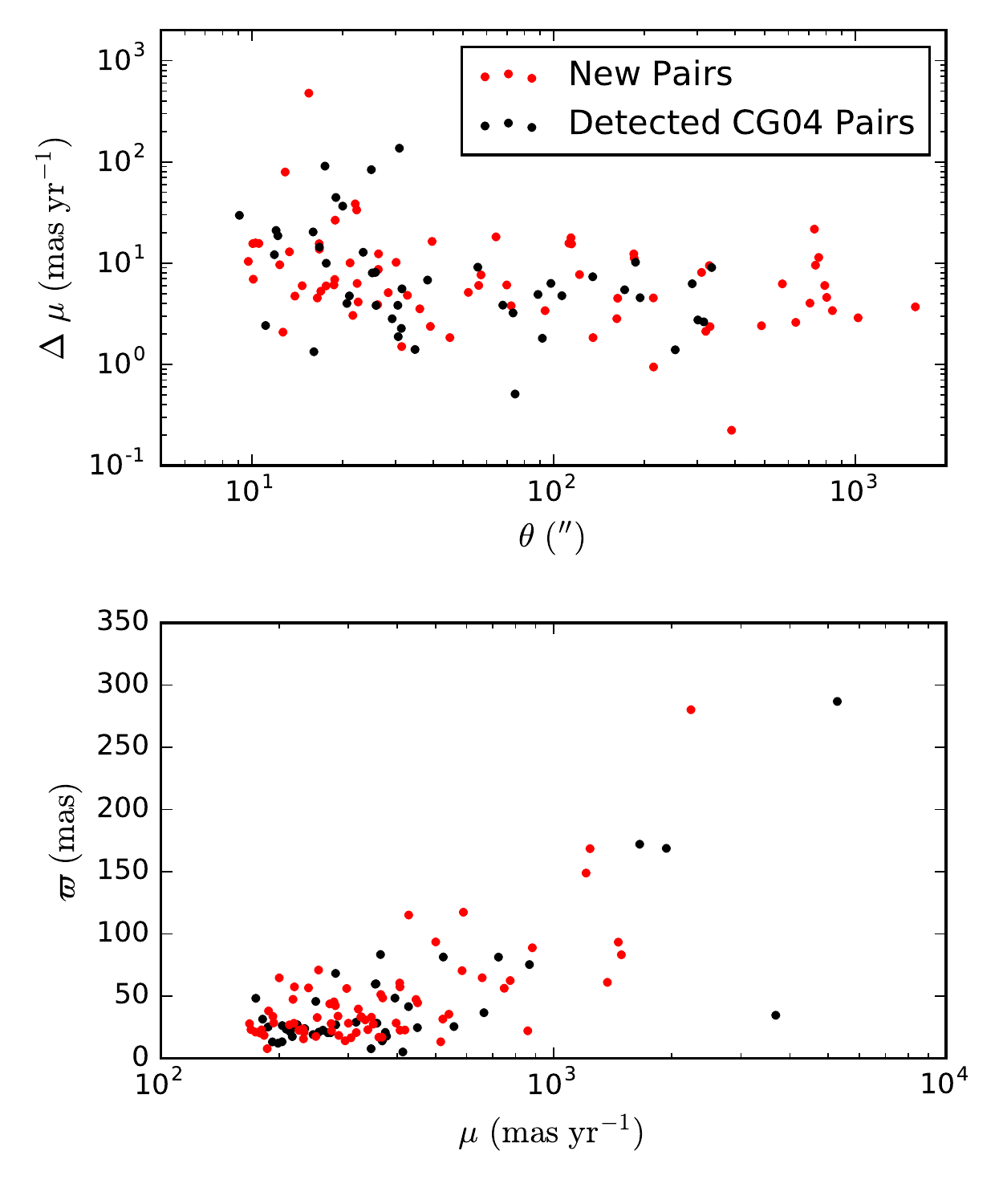}
\caption{The 44 wide binaries with {\it Hipparcos} parallaxes in the CG04 catalog (black), all of which we recover, and the 72 newly detected wide binaries (red). {\it Top} --- Proper motion difference as a function of angular separation for pairs in the two samples. {\it Bottom} --- Parallax as a function of proper motion. On average, our newly detected binaries are found at larger angular separations, and have smaller parallaxes and proper motions than the previously identified rNLTT pairs with parallaxes. }
\label{fig:rNLTT_detections}
\end{center}
\end{figure}

\citet[][]{chaname04} found 999 wide binaries in the rNLTT catalog; 44 of these have {\it Hipparcos} parallaxes for both components. There are some important differences between our search algorithm and the approach of CG04. CG04 made proper motion cuts, using a limit for $\Delta \mu$ of 20 mas yr$^{-1}$, a factor of three larger than typical uncertainties on $\Delta \mu$. Our method includes uncertainties in proper motion as well as a quantification of binary probability as a function of $\theta$, $\Delta \mu$, and position in position-proper motion phase space. On the other hand, to remove potential contaminants, CG04 calculated a modified reduced proper motion to discern between disk and halo populations. Clearly, true binaries cannot contain stars from different populations, a powerful constraint that is not built into our method (matching TGAS parallaxes does provide a similar, alternative constraint).

Given these differences, we did not expect the resulting binary samples to be identical. Encouragingly, however, we recovered all 44 CG04 pairs with {\it Hipparcos} parallaxes; the other 72 pairs in our sample are newly identified wide binaries. We compare the two samples in Figure~\ref{fig:rNLTT_detections}.

We leave discussion and follow-up of these pairs for future work. For now we note that even in catalogs such as rNLTT that have been mined for wide binaries, there may be a substantial number of true pairs still to be identified. 

\bsp	
\label{lastpage}
\end{document}